\documentclass[prb,twocolumn,amsmath,amssymb,floatfix,footinbib,bibnotes,longbibliography]{revtex4-2}
\pdfoutput=1

\usepackage{hyperref}
\hypersetup{colorlinks=true,citecolor=blue,linkcolor=magenta}

\usepackage{xcolor}

\usepackage{amsmath}
\usepackage{graphicx}
\usepackage{changepage}
\usepackage{fancyhdr}
\usepackage{amsthm, amssymb}
\usepackage{breqn} % Helps with breaking long equations
\usepackage{parskip} % Adds spacing between paragraphs for readability
\usepackage{float}

\usepackage{array}
\usepackage{mathtools}
\usepackage[normalem]{ulem}
\usepackage{physics}

\newcommand{\beq}{\begin{eqnarray}}
\newcommand{\eeq}{\end{eqnarray}}

\def \titlename {Schwinger-Keldysh field theory for operator R\'{e}nyi entropy and entanglement growth in non-interacting systems with sub-ballistic transports}
\def \authornames{Priesh Roy, and Sumilan Banerjee}
\def \affiliations{Centre for Condensed Matter Theory, Department of Physics, Indian Institute 
	of Science, Bangalore 560012, India}

\newcommand{\dif}{\mathop{}\,\mathrm{d}}

\newcommand{\olsi}[1]{\,\overline{\!{#1}}} % overline short italic

\newcommand{\ops}{\ensuremath{{{\mathcal{O}}}}}

 \newcommand{\bcen}{\begin{center}}
 \newcommand{\ecen}{\end{center}}
 \newcommand{\btab}{\begin{tabular}}
 \newcommand{\etab}{\end{tabular}}
 \newcommand{\bdes}{\begin{description}}
 \newcommand{\edes}{\end{description}}

\newcommand{\balgn}{\begin{align}}
\newcommand{\ealgn}{\end{align}}

 \newcommand{\bary}{\begin{array}}
 \newcommand{\eary}{\end{array}}
 \newcommand{\benum}{\begin{enumerate}}
 \newcommand{\eenum}{\end{enumerate}}
 \newcommand{\bitem}{\begin{itemize}}
 \newcommand{\eitem}{\end{itemize}}
 \newcommand{\bmat}{\ensuremath{\begin{pmatrix}}}
 \newcommand{\emat}{\ensuremath{\end{pmatrix}}}
 \newcommand{\lf}{\left (}
 \newcommand{\rf}{\right )}

 \newcommand{\lt}{\left [}
 \newcommand{\rt}{\right ]}

\begin{document}
	
	\title{\titlename}
	\author{\authornames}
	\affiliation{\affiliations}
	\email{prieshroy@iisc.ac.in, sumilan@iisc.ac.in}
	% \date\today

\begin{abstract}
The notion of operator growth in quantum systems furnishes a bridge between transport and the generation of entanglement between different parts of the system under quantum dynamics. We define a measure of operator growth in terms of subsystem operator R\'{e}nyi entropy, which provides a state-independent measure of operator growth, unlike entanglement entropies, and the usual measures of operator growth like out-of-time-order correlators. We show that the subsystem operator R\'{e}nyi entropy encodes both spatial and temporal information, and thus can directly connect to transport for a local operator related to a conserved quantity. We construct a unified Schwinger-Keldysh (SK) field theory formalism for the time evolution of operator R\'{e}nyi entropy and entanglement entropies of initial pure states. We use the SK field theory to obtain the operator R\'{e}nyi and state entanglement entropies in terms of infinite-temperature and vacuum Keldysh Green's functions, respectively, for non-interacting systems. We apply the method to explore the connection between operator and entanglement growth, and transport in non-interacting systems with quasiperiodic and random disorder, like the one- and two-dimensional Aubry-Andr\'{e} models and the two-dimensional Anderson model. In particular, we show that the growth of subsystem operator R\'{e}nyi entropy and state von Neumann and R\'{e}nyi entanglement entropies can capture both ballistic and sub-ballistic transport behaviors, like diffusive and anomalous diffusive transport, as well as localization in these systems.
\end{abstract}

\maketitle 	

%%%%%%%%%%%%%%%%%%%%%%%%%%%%%%%%%%%%%%%%%%%%%%%%%%%%%%%%%%%%%%%
\section{Introduction} \label{sec:Intro}
The investigation of quantum information propagation, for example, characterized by growth of operators and entanglement of quantum states under quantum many-body time evolution, has emerged as a pivotal research area in condensed matter and high-energy physics \cite{Lieb1972,AbaninAltmanRMP,FisherCircuitReview,Sekino2008,Maldacena2016}. In particular, it is an exciting avenue to study the connections between diagnostics of quantum information propagation and conventional characterizations of quantum many-body systems, such as in terms of transport of conserved quantities. Quantum information propagation is commonly quantified in terms of out-of-time-order correlators (OTOC) \cite{LarkinOvchinikov}, and operator and state entanglement entropies \cite{Bandyopadhyay2005,Prosen2007,AbaninAltmanRMP,FisherCircuitReview}. For example, entanglement of a pure state with reduced density matrix $\rho_A$ of a subsystem is quantified in terms of $n$-th R\'{e}nyi entanglement entropies, $S_A^{(n)}\equiv [1/(1-n)]\ln{\mathrm{Tr}(\rho_A^n)}$, like the second R\'{e}nyi entropy $S_A^{(2)}$ and ($n\to 1$) von Neumann entropy $S_A=S_A^{(n\to 1)}$. Starting with unentangled initial states, in generic interacting thermalizing systems with diffusive transport of conserved quantities, the R\'{e}nyi entanglement entropies $S_A^{(n>1)}(t)$ are found to grow diffusively as $\sim \sqrt{t}$ with time $t$, while the $S_A(t)$ follows ballistic growth $\sim t$ \cite{Kim2013,PollMann,Ludwig}. In contrast, all R\'{e}nyi entropies and the von Neumann entropy are expected to grow ballistically in clean integrable systems \cite{Calabrese2005,Calabrese2007,Alba2017,Alba2018,Fagotti2008,Alba2017_a,Mestyan2018,Alba2021} with ballistic transport, as well as in systems without any conservation laws \cite{Keyserlingk2018,Nahum2018,Zhou2019,Bertini2019}.

The above correspondence between transport and state entanglement growth has been rationalized based on the time dependence of entanglement eigenvalues or Schmidt values of the reduced density matrix in numerical studies of integrable and non-integrable many-body models \cite{PollMann,Rakovszky2019}. Analytical bounds \cite{Ludwig} and calculations in unitary quantum circuits with or without conservation laws \cite{Keyserlingk2018,Nahum2018,FisherCircuitReview} also support the numerical results. In an interacting diffusive system, the maximum entanglement eigenvalue $\Lambda_{max}$, which dominates $S_A^{(n>1)}(t)$ in the long time limit, is found to decay \emph{diffusively} as $\ln\Lambda_{max}(t)\sim -\sqrt{t}$. This is in contrast to much faster decay of the typical Schmidt values, $\ln\Lambda_{typ}\sim -t$, which control the growth of $S_A(t)$. In ballistic integrable systems and systems without conservation laws, the slow decay of $\Lambda_{max}$ is absent, and it decays exponentially in time, like all other entanglement eigenvalues. It is an interesting question how the above scenario gets modified in systems that exhibit sub-ballistic transport, but anomalous diffusion, like sub or super-diffusive behaviors. The latter can be observed in certain integrable systems \cite{Vasseur}, whereas signatures of sub-diffusive transport are often observed in low-dimensional interacting systems, e.g., in the ergodic phase of interacting one-dimensional (1D) random and quasiperiodic systems close to the many-body localization (MBL) transition \cite{Agarwal,Agarwal2017,Pixley,Yoo2020}. However, calculations of entanglement and operator growth in such interacting systems are typically limited to numerical computations for relatively small systems. In this work, we study one- and higher- dimensional non-interacting quasiperiodic and random systems where transport characteristics can be tuned over the full range, from ballistic to sub-diffusive. In the non-interacting systems, the time-evolution of operator and entanglement growth can be computed efficiently for sufficiently large systems.

To this end, we propose a diagnostic of operator growth in terms of a subsystem operator second R\'{e}nyi entropy $S_{\mathcal{O}A}^{(2)}(t)$. The quantity is defined through partial trace over a subsystem $B$, $\mathrm{Tr}_B[\mathcal{O}(t)]$, of a local operator, time evolved through the system Hamiltonian $\mathcal{H}$, $\mathcal{O}(t)=e^{\mathrm{i}\mathcal{H}t}\mathcal{O}e^{-\mathrm{i}\mathcal{H}t}$. The operator R\'{e}nyi entropy $S_{\mathcal{O}A}^{(2)}(t)$ provides a state-independent measure of operator growth and is defined in a spirit similar to the recently studied partial spectral form factor \cite{Joshi}. However, in contrast to the latter, $S_{\mathcal{O}A}^{(2)}(t)$ contains spatial information in addition to the information of the many-body energy spectrum and eigenfunctions.  The subsystem operator R\'{e}nyi entropy $S_{\mathcal{O}A}^{(2)}(t)$ is distinct from usual operator space entanglement entropy \cite{Zanardi,Prosen2007,Luitz, Dubail} which requires operator-to-state mapping through the construction of a doubled Hilbert space. We show that $S_{\mathcal{O}A}^{(2)}(t)$ diagnoses characteristics of ballistic and sub-ballistic transport, e.g., sub- and super-diffusive transport as well as lack of transport or localization, in 1D and 2D quasiperiodic Aubry-Andr'{e} models \cite{AA, Devakul,Sutradhar} across delocalization-localization transitions. The operator R\'{e}nyi entropy also captures diffusive behavior in a finite-sized 2D Anderson model with random disorder where the weak localization correction has not set in. We also compare the results for operator R\'{e}nyi entropy $S_{\mathcal{O}A}^{(2)}(t)$ with the time evolution of usual state second R\'{e}nyi ($S_A^{(2)}$) and von Neumann ($S_A$) entanglement entropies starting from an unentangled pure product state. We show that the state entanglement growth is also able to capture the transport features, though the operator R\'{e}nyi entropy provides clearer diagnostics of the nature of transport due to the explicit dependence on the spatial location of the local operator involved. 

The growth of operator and state entropies in the ballistic and sub-ballistic regimes can be rationalized in terms of the decay of the Schmidt values, as in the interacting models \cite{PollMann,Rakovszky2019,Ludwig}. However, we show that, unlike interacting systems, there is no distinction between the decays of maximum and typical Schmidt values. All the Schmidt values in these non-interacting systems exhibit either ballistic or sub-ballistic decay following the transport behavior. This is expected for non-interacting systems \cite{Ludwig}, like 1D and 2D quasiperiodic Aubry-Andr'{e} models, where single-particle wave functions of individual particles spread ballistically or sub-ballistically \cite{Purkayastha,Sutradhar}.

To compute state entanglement entropies starting from a pure product state, we employ a recently developed unified Schwinger-Keldysh (SK) non-equilibrium field theory framework following Refs.~\onlinecite{Haldar,Bera,Perugu}. We also extend the SK field theory formalism for operator second R\'{e}nyi entropy. Though the SK field theory formalism is general and can be applied to interacting systems within suitable approximations \cite{Bera} or for solvable models \cite{Haldar,Perugu}, in this work, we apply this method to non-interacting random and quasiperiodic systems. To this end, we obtain expressions for operator and state entropies in terms of single-particle infinite-temperature and vacuum Green's functions of the non-interacting systems. As a result, these expressions can be easily evaluated for reasonably large systems to obtain $S_{\mathcal{O}A}^{(2)}(t)$, $S_A^{(2)}(t)$ and $S_A(t)$ at arbitrary time $t$.

The rest of the paper is organized as follows. In Sec.~\ref{sec:Definition}, we define the operator R\'{e}nyi entropy for a subsystem and state entanglement entropies. Sec.~\ref{sec:PathIntegral} describes the Schwinger-Keldysh (SK) coherent-state path integral formalism for the operator second R\'{e}nyi entropy and the second R\'{e}nyi entanglement entropy for an initial arbitrary occupation basis product state. Numerical results obtained from the applications of the SK field theory to non-interacting systems, namely 1D and 2D quasiperiodic Aubry-Andr\'{e} models and the 2D random Anderson model, with transport varying from ballistic to various sub-ballistic behaviors are discussed in Sec.~\ref{sec:Results}. In Sec.~\ref{sec:Conclusion}, we conclude and discuss some future directions. The details of some of the analytical derivations and additional numerical results are provided in the Appendices.

\begin{figure}[h]
    \centering
    \includegraphics[width=0.65\linewidth]{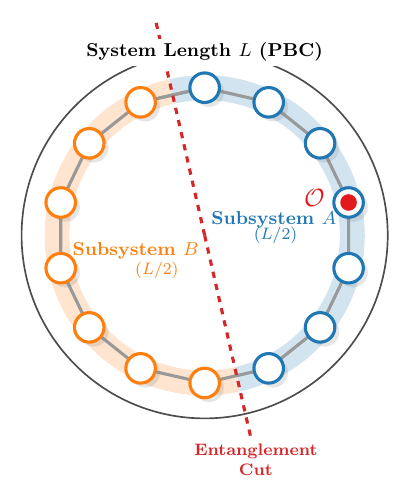}
    \caption{Schematic of the division of the system into two subsystems, e.g., in one dimension, for the calculation of the operator R\'{e}nyi entropy and state entanglement entropies. A system of length $L$ with periodic boundary condition, is partitioned into two halves $A$ and $B$. The operator $\mathcal{O}$ is placed at a distance $\sim L/4$ from the boundaries between $A$ and $B$.}
    \label{fig:schematic1}
\end{figure}

\section{Operator R\'{e}nyi entropy and state entanglement entropies} \label{sec:Definition}
\subsection{Subsystem Operator R\'{e}nyi entropy}
We consider a time-evolved Hermitian operator, i.e., $\mathcal{O}(t)=e^{\mathrm{i}\mathcal{H}t}\mathcal{O}e^{-\mathrm{i}\mathcal{H}t}$ and define a partial trace over subsystem $B$, $\rho_{\mathcal{O}A}(t)=\mathrm{Tr}_B[\mathcal{O}(t)]/Z_{\mathcal{O}}$, as shown in Fig.~\ref{fig:schematic1}. Here $Z_{\mathcal{O}}=\mathrm{Tr}[\mathcal{O}]$ is a normalization. For convenience in the analytical formulation below, we consider the trace (Tr) to be over the full Fock space, not restricted to a fixed particle number sector, even though the Hamiltonian $\mathcal{H}$ and operator $\mathcal{O}$, we consider, conserve particle number. We use $\rho_{\mathcal{O}A}(t)$ to define a subsystem operator R\'{e}nyi entropy $S_{\mathcal{O}A}^{(2)}(t)$,
\begin{align} \label{eq:DefOperatorRenyi}
    e^{-S_{\mathcal{O}A}^{(2)}(t)} &= \Tr_A{\lt \rho_{\ops A}^2(t)\rt}.
\end{align}
At $t = 0$, $S_{\mathcal{O} A}^{(2)} \neq 0$ and is given by a system-size-dependent constant factor, e.g., $\lf L/2 - 1 \rf \ln{2}$ for a half subsystem of length $L/2$, if $\mathcal{O}$ is localized initially in subsystem $A$, as shown in Appendix \ref{app:S0}. So, to quantify operator growth, we calculate $\Delta S_{\mathcal{O} A}^{(2)}(t) = S_{\mathcal{O} A}^{(2)}(t) - S_{\mathcal{O} A}^{(2)}(0) $.

More insights into the content of $S_{\mathcal{O} A}^{(2)}(t)$ can be obtained from the representation of the operator, $\mathcal{O}(t)=\sum_{nm}\mathcal{O}_{nm}e^{\mathrm{i}(E_n-E_m)t}|E_n\rangle\langle E_m|$, in terms of the many-body eigenenergies $E_n$ and eigenstates $|E_n\rangle$ of $\mathcal{H}$. This leads to
\begin{align}
e^{-S_{\mathcal{O}A}^{(2)}(t)}&= Z_{\mathcal{O}}^{-2} \sum_{nmn'm'} \mathcal{O}_{nm} \mathcal{O}_{n'm'} e^{\mathrm{i}(E_n-E_m+E_{n'}-E_{m'})t}\nonumber \\
&\times \mathrm{Tr}_A\left[\mathrm{Tr}_B\left(|E_n\rangle \langle E_m|\right)\mathrm{Tr}_B\left(|E_{n'}\rangle \langle E_{m'}|\right)\right],
\end{align}
where $\mathcal{O}_{nm}=\langle E_n|\mathcal{O}|E_m\rangle$ is the matrix element of the operator. As evident from above, the subsystem operator R\'{e}nyi entropy contains information about the many-body eigenstates, eigenenergies, as well as the operator. The latter encodes the spatial information if $\mathcal{O}$ is a local operator and thus can provide a suitable state-independent probe of the interplay of transport and operator growth.

\subsection{State Entanglement}
We also compute the time evolution of the second R\'{e}nyi and von Neumann entanglement entropies, $S_A^{(2)}(t)$ and $S_A(t)$, for a pure state $|\psi(0)\rangle$ from the reduced density matrix of the subsystem $A$, $\rho_A(t)=\mathrm{Tr}_B\rho(t)$, where $\rho(t) = e^{-\mathrm{i}\mathcal{H}t} |\psi(0)\rangle \langle\psi(0)| e^{\mathrm{i}\mathcal{H}t}$, e.g.,
\begin{align}
  e^{-S_A^{(2)}(t)} = \Tr_A{\lt \rho^2_A(t) \rt}. 
\end{align}
Since the Hamiltonians considered in this work conserve total particle number, we restrict the state $|\psi(0)\rangle$ in a fixed particle number sector. In particular, for the non-interacting spinless fermionic Hamiltonians used in this work, we consider the half-filled particle number sector for the computation of the state entanglement entropies. As a result, all the traces that appear in the formulation for the state entanglement entropies below are also restricted to the half-filled particle number sector for the system, unlike the traces over all the particle-number sectors used for operator R\'{e}nyi entropy in the preceding section.
%%%%%%%%%%%%%%%%%%%%%%%%%%%%%%%%%%%%%%%%%%%%%%%%%%%%%%%%%%%%%%%

\section{Schwinger-Keldysh (SK) field theory for operator R\'{e}nyi and state entanglement entropies}\label{sec:PathIntegral}

We develop a unified SK path integral representation, e.g., for a quantity of the form $\mathrm{Tr}_A[(\mathrm{Tr}_BO)^2]$ [$O=\mathcal{O}(t)/Z_{\mathcal{O}},~\rho(t)$], to compute $S_{\mathcal{O}A}^{(2)}(t)$ and $S_A^{(2)}(t)$. The complexity of this task stems from the requirement to link distinct replicas of the coherent-state path integral via suitable boundary conditions on the fields \cite{cardyCalabrese,Calabrese2009,Casini2009,Chakraborty_2021}, e.g., the Grassmann fields for the fermionic systems that we consider here. Following Ref.~\onlinecite{Haldar}, we use the identity
\begin{align} \label{eq:TraceIdentity}
\mathrm{Tr}_A\left[\left( \mathrm{Tr}_B O\right)^2\right]&=\int  f_A(\xi_1, \xi_2) \prod_\alpha d^2\xi_\alpha \Tr \lt O D_A(\xi_\alpha)\rt 
\end{align}
in terms of fermionic displacement operators \cite{CahillGlauber},
\begin{equation}
    D_A(\xi_\alpha)=e^{\sum_{i\in A} c_i^\dagger \xi_{\alpha i}}e^{-\sum_{i\in A}\bar{\xi}_{\alpha i}c_i},
\end{equation}
where $\xi_\alpha \equiv \{\olsi{\xi}_{\alpha i} , \xi_{\alpha i}\}$ denotes sets of auxiliary Grassmann variables at site $i\in A$ with $L_A$ sites (out of total $L$ sites) on two \emph{entanglement replicas} $\alpha=1,2$, and $\dif^2\xi_\alpha = \prod_{ i\in A} d \olsi{\xi}_{\alpha i} d \xi_{\alpha i}$. The Gaussian function $f_A(\xi_1,\xi_2)$, that couples the two traces in Eq.\eqref{eq:TraceIdentity}, is given by
\begin{equation}
f_A(\xi_1, \xi_2) = 2^{L_A}e^{ -(1/2) \sum_{i\in A} \lf \olsi{\xi}_{1i} \xi_{1i} + \olsi{\xi}_{2i} \xi_{2i} + \olsi{\xi}_{1i} \xi_{2i} - \olsi{\xi}_{2i} \xi_{1i} \rf}.
\end{equation}
The above identity [Eq.\eqref{eq:TraceIdentity}] allows us to write a path integral for $\mathrm{Tr}[OD_A(\xi_\alpha)]$ and eliminate the complication of evaluating partial traces, as we discuss below.

%%%%%%%%%%%%%
\subsection{The SK field theory for subsystem operator R\'{e}nyi entropy}\label{sec:PathIntegralPSFF}
\begin{figure}[H]
\centering
\includegraphics[width=0.8\linewidth]{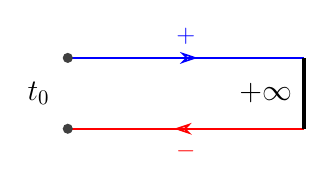}
\caption{The Schwinger-Keldysh closed-time contour for an arbitrary initial density matrix $\rho_0$, extending from $t_0$ to $+\infty$ and returning to $t_0$. The two branches represent the forward ($+$) and backward ($-$) time evolutions, respectively.}
\label{fig:ctc}
\end{figure}

Using the identity of Eq.\eqref{eq:TraceIdentity}, from Eq.\eqref{eq:DefOperatorRenyi}, we write 
\begin{align} \label{eq:OpRenyi_WignerFn}
e^{-S_{\ops A}^{(2)}(t)}&=\frac{1}{Z_{\ops}^2}\int f_A(\xi_1,\xi_2)\prod_\alpha d^2\xi_\alpha\chi_{\mathcal{O}}(\xi_\alpha,t),
\end{align}
where $\chi_{\mathcal{O}}(\xi,t)$ is the fermionic Wigner characteristic function \cite{Haldar,Moitra2020} for the operator $\mathcal{O}(t)$,
\begin{align}
\chi_{\ops}(\xi_\alpha,t)&=\mathrm{Tr}\left[e^{-\mathrm{i}\mathcal{H}t}D_A(\xi_\alpha)e^{\mathrm{i}\mathcal{H}t} \mathcal{O}\right].
\end{align}
The above can be written as a SK path integral on a \emph{reverse} Keldysh contour \cite{Kamenev} [$(0-,-\infty-)\cup (-\infty+,0+)$], with backward evolution ($e^{\mathrm{i}\mathcal{H}t}$, $``-"$ branch) followed by a forward ($e^{-\mathrm{i}\mathcal{H}t}$, $``+"$ branch) evolution. However, since we only consider time-reversal invariant systems and operators here, we can write the path integral on the usual Keldysh contour \cite{Kamenev} ($\mathcal{C}=(0+,\infty+)\cup (\infty-,0-)$) [Fig. \ref{fig:ctc}] for the convenience of notation, and to maintain the same contour for operator and state entropies (see later). Effectively, we consider the operator evolution over a time $-t$ with $t>0$, i.e. we redefine $\mathcal{O}(t)\equiv e^{-\mathrm{i}\mathcal{H}t}\ops e^{\mathrm{i}\mathcal{H}t}$. As a result,
\begin{align} \label{eq:OpWignerFn_1}
\chi_{\mathcal{O}}(\xi_\alpha,t)=\int \mathcal{D}(\bar{c},c)e^{\mathrm{i}S[\bar{c},c,\xi_\alpha]}\langle c(0+)|\mathcal{O}|-c(0-)\rangle,
\end{align}
where $\{\bar{c}_i(t\pm),c_i(t\pm)\}$ are Grassmann fields on the Keldysh contour $z=t\pm\in \mathcal{C}$ and 
\begin{align} \label{eq:OpAction_1}
&S[\bar{c},c,\xi_\alpha]=\int_{{\mathcal{C}}}dz \left[\sum_i \bar{c}_i(z)\mathrm{i}\partial_zc_i(z)-\mathcal{H}(\bar{c},c)\right]  \\
&-\mathrm{i}\int_{{\mathcal{C}}}dz\sum_{i\in A}\left[\bar{c}_i(z)\delta_{\mathcal{C}}(z,t^++)\xi_{\alpha i}-\bar{\xi}_{\alpha i}\delta_{\mathcal{C}}(z,t+)c_i(z)\right] \nonumber
\end{align}
is the action where the insertion of the displacement operator at time $t$ leads to an additional \emph{kick} term \cite{Haldar} in the second line above.
%%%%%%%%%%%%%%%

%%%%%%%%%%%%%%%%%%%%%%
In Eq.\eqref{eq:OpWignerFn_1}, the information about the operator is encoded in its coherent-state matrix element $\langle c(0+)|\ops|-c(0-)\rangle$. In this work, to connect operator growth with the transport of a conserved quantity, namely the fermion number, we consider the number operator $\ops=\hat{n}_{l}=c_{l}^\dagger c_{l}$ at a specific site $l$ (see Fig.~\ref{fig:schematic1}). We exponentiate the matrix element by introducing additional fermionic source fields to include the matrix element within the action. To this end, we write
\begin{align} \label{eq:OpMatrixElement}
&\langle c(0+)|\hat{n}_l|-c(0-)\rangle =\left[-\bar{c}_l(0+)c_l(0-)\right]e^{-\sum_i \bar{c}_i(0+)c_i(0-)}\nonumber \\
~&=\int d^2\zeta_l e^{\bar{c}_l(0+)\zeta_l-\bar{\zeta}_lc_l(0-)}e^{-\sum_i \bar{c}_i(0+)c_i(0-)}.
\end{align}
Using the above in Eq.\eqref{eq:OpWignerFn_1}, we obtain
\begin{align}
\chi_{\ops}(\xi_\alpha,t)=\int d^2\zeta_{ l}\mathcal{D}(\bar{c},c)e^{\mathrm{i}S[\bar{c},c,\xi_\alpha,\zeta_{l}]},
\end{align}
where $d^2\zeta_{ l}=d\bar{\zeta}_{ l}d\zeta_{l}$. In Eq.\eqref{eq:OpMatrixElement}, the term $\exp[-\sum_i \bar{c}_i(0+)c_i(0-)]$  acts as an \emph{infinite-temperature} boundary condition to the SK path integral. For the non-interacting systems considered here, using Eq.\eqref{eq:OpAction_1}, we can write the action appearing above as
\begin{align} \label{eq:OpAction_NonInt}
&S[\bar{c},c,\xi_\alpha,\zeta_{ l}]=\int_{{\mathcal{C}}}dz_1 dz_2 \sum_{ij} \bar{c}_i(z)G_{ij}^{-1}(z_1,z_2)c_j(z_2) \nonumber \\
&-\mathrm{i}\int_{\mathcal{C}}dz\sum_{i}
\left\{\bar{c}_i(z)\left[\delta_{\mathcal{C}}(z,t^++)\delta_{i\in A}\xi_{\alpha i}+\delta_{\mathcal{C}}(z,0+)\delta_{il}\zeta_{l}\right] \right.\nonumber \\
&\left.-\left[\bar{\xi}_{\alpha i}\delta_{\mathcal{C}}(z,t+)\delta_{i\in A}+\bar{\zeta}_{l}\delta_{\mathcal{C}}(z,0-)\delta_{il}\right]c_i(z)\right\},
\end{align}
where $G_{ij}(z_1,z_2)$ is the non-interacting infinite-temperature Green's function on the Keldysh contour [Fig. \ref{fig:ctc}], as discussed in Appendix \ref{app:NIGF}. $\delta_{i\in A}=1$ for $i\in A$ and zero otherwise.
%%%%%%%%%%%%

%%%%%%%%%%%%%%%%%%%%%%%%%%%%%
Since the action [Eq.\eqref{eq:OpAction_NonInt}] is quadratic, we can integrate out the dynamical Grassmann fields $\{\bar{c}_i(z),c_{i}(z)\}$, and subsequently the auxiliary fields $(\bar{\zeta}_{l},\zeta_{l})$, to obtain an expression for the Wigner function of Eq.\eqref{eq:OpWignerFn_1},
\begin{align} \label{eq:OpWignerFn_3}
\chi_{\mathcal{O}}(\xi_\alpha,t)&=\mathrm{det}\left[-\mathrm{i}G^{-1}\right]\left[\mathrm{i}G_{ll}(0-,0+)\right]e^{\sum_{ij\in A}\bar{\xi}_{\alpha i}C_{\ops,ij}(t)\xi_{\alpha j}}.
\end{align}
Here 
\begin{align} \label{eq:OpCorrMatrix}
&C_{\ops,ij}(t)=-\mathrm{i}G_{ij}(t+,t^++) \nonumber \\
&~+\mathrm{i}G_{il}(t+,0+)\left[G_{ll}(0-,0+)\right]^{-1}G_{lj}(0-,t^++)
\end{align}
are the elements of a $L_A\times L_A$ correlation matrix $\mathbb{C}_\ops$. The Green's functions appearing above can be identified with usual \emph{lesser and greater} Green's functions on the Keldysh contour \cite{Kamenev}, namely, for $t>0$, $G_{ij}(t+,t^++)=G^{<}(t,t)$, $G_{il}(t+,0+)=G^{>}(t,0)$, $G_{ll}(0-,0+)=G^{>}(0,0)$ and $G_{lj}(0-,t^++)=G_{lj}^{>}(0,t)$. These Green's functions can be easily evaluated in terms of single-particle eigenstates and eigenenergies for non-interacting systems (see Appendix \ref{app:NIGF}). Using a similar SK path integral construction, to the one above for $\chi_\ops(\xi_\alpha,t)$ [Eq.\eqref{eq:OpWignerFn_1}], it can be easily shown that $Z_\ops=\mathrm{Tr}[e^{\mathrm{i}\mathcal{H}t}e^{-\mathrm{i}\mathcal{H}t}\mathcal{O}]=\mathrm{det}\left[-\mathrm{i}G^{-1}\right]\left[\mathrm{i}G_{ll}(0-,0+)\right]$. As a result, using Eq.\eqref{eq:OpWignerFn_3} in Eq.\eqref{eq:OpRenyi_WignerFn} and integrating out the auxiliary fields $\{\bar{\xi}_{\alpha i},\xi_{\alpha i}\}$, we finally obtain
\begin{align}
S_{\ops A}^{(2)}(t)&=-\mathrm{tr}\ln\left[\left(\mathbb{I}-\mathbb{C}_\ops(t)\right)^2+\mathbb{C}_\ops^2(t)\right] \label{eq:OperatorRenyi}
\end{align}
for the R\'{e}nyi entropy for the operator $n_l$ for the $A$ subsystem. Here $\mathbb{I}$ is the identity matrix. Thus we can obtain the entire time dependence of $S_{\ops A}^{(2)}(t)$ in terms of the infinite-temperature single-particle Green's functions (Appendix \ref{app:NIGF}).

We benchmark the expression for $S_{n_l A}^{(2)}(t)$ in Eq.\eqref{eq:OperatorRenyi}, evaluated in terms of single-particle Green's functions [Eq.\eqref{eq:OpCorrMatrix}], with $S_{n_l A}^{(2)}(t)$ obtained from the full many-body exact diagonalization (ED) of 1D tight-binding chain with $L=8$ sites, as discussed in Appendix \ref{app:NIGF}.
%%%%%%%%%%%%%%%%%%%%%%%%%%%%%

%%%%%%%%%%%%%%%%%%%%%%%%%%%%%%%%%
\subsection{The SK field theory for the growth of R\'{e}nyi entanglement entropy of initial pure product states}\label{sec:PathIntegralSFF}

We consider initial pure product states, $|\psi(0)\rangle=|n_1n_2\cdots n_L\rangle\equiv |n\rangle$, which are basis states formed from real-space occupation of sites $i=1,\cdots,L$. Similar to operator R\'{e}nyi entropy in the preceding section, using $O=\rho(t)=e^{-\mathrm{i}\mathcal{H}t}|n\rangle \langle n|e^{\mathrm{i}\mathcal{H}t}$ in Eq.\eqref{eq:TraceIdentity}, we obtain for the second R\'{e}nyi entanglement entropy of the subsystem $A$ as,
\begin{align}
e^{-S_A^{(2)}(t)}&=\int f_A(\xi_1,\xi_2)\prod_\alpha d^2\xi_\alpha\chi(\xi_\alpha,t),
\end{align}
where the Wigner function, $\chi(\xi_\alpha,t)=\mathrm{Tr}[\rho(t)D_A(\xi_\alpha)]$ for $\rho(t)$ is given by
\begin{align} \label{eq:WignerFn_1}
\chi(\xi_\alpha,t)=\int \mathcal{D}(\bar{c},c)e^{\mathrm{i}S[\bar{c},c,\xi_\alpha]}\langle c(0+)|n\rangle\langle n|-c(0-)\rangle.
\end{align}
In the above, $S[\bar{c},c,\xi_\alpha]$ is the same action that appears in Eq.\eqref{eq:OpAction_1}.
Following Ref.~\onlinecite{Perugu}, using the identities $\langle n|-c\rangle=c_L^{n_L}\cdots c_1^{n_1}$, $\langle c|n\rangle=(-\bar{c}_1)^{n_1}\cdots (-\bar{c}_L)^{n_L}$ and $(-\bar{c}_i(0+)c_i(0-))=\int d^2\zeta_i \exp[\bar{c}_i(0+)\zeta_i-\bar{\zeta}_ic_i(0-)]$, we exponentiate the matrix element,
\begin{align}
\langle c(0+)|n\rangle \langle n|-c(0-)\rangle&=\int \prod_{i\in I}d^2\zeta_i e^{\sum_{i\in I}[\bar{c}_i(0+)\zeta_i-\bar{\zeta}_ic_i(0-)]},
\end{align}
where $I$ denotes the set of initially occupied sites, i.e., $n_i=1$ for $i\in I$ and $n_i=0$ otherwise. A crucial difference of the above matrix element from Eq.\eqref{eq:OpMatrixElement} is the absence of the term $\exp[-\sum_i \bar{c}_i(0+)c_i(0-)]$.

Thus, we obtain 
\begin{align} \label{eq:WignerFn_2}
\chi(\xi_\alpha,t)=\int \prod_{i\in I}d^2\zeta_i \mathcal{D}(\bar{c},c)e^{\mathrm{i}S[\bar{c},c,\xi_\alpha,\zeta]}.
\end{align}
Again, for the non-interacting systems, the action is given by
\begin{align} \label{eq:OpAction_NonInt}
&S[\bar{c},c,\xi_\alpha,\zeta_i ]=\int_{{\mathcal{C}}}dz_1 dz_2 \sum_{ij} \bar{c}_i(z)G_{0,ij}^{-1}(z_1,z_2)c_j(z_2) \nonumber \\
&-\mathrm{i}\int_{\mathcal{C}}dz\sum_{i}
\left\{\bar{c}_i(z)\left[\delta_{\mathcal{C}}(z,t^++)\delta_{i\in A}\xi_{\alpha i}+\delta_{\mathcal{C}}(z,0+)\delta_{i\in I}\zeta_{i}\right] \right.\nonumber \\
&\left.-\left[\bar{\xi}_{\alpha i}\delta_{\mathcal{C}}(z,t+)\delta_{i\in A}+\bar{\zeta}_{i}\delta_{\mathcal{C}}(z,0-)\delta_{i\in I}\right]c_i(z)\right\},
\end{align}
where $G_{0,ij}(z_1,z_2)$ is the non-interacting \emph{vacuum} Green's function on the Keldysh contour [Fig. \ref{fig:ctc}], as discussed in Appendix \ref{app:NIGF}. $\delta_{i\in I}=1$ for $i\in I$ and zero otherwise.

Following procedure similar to the operator R\'{e}nyi entropy in the preceding section, we finally obtain for the second state R\'{e}nyi entanglement entropy,
\begin{equation}\label{eq:StateRenyiEE}
S_A^{(2)}(t) = - \mathrm{tr}\ln\left[ \left(\mathbb{I} - \mathbb{C}(t) \right)^2 + \mathbb{C}^2(t) \right].
\end{equation}
In this case, the correlation-matrix $\mathbb{C}(t)$ is defined through the elements
\begin{align}
\label{eq:StateCorrMatrix}
&C_{ij}(t) = -\mathrm{i} G_{0,ij}(t+, t^++) \notag \\
&~\quad + \mathrm{i} \sum_{kl \in I} G_{0,ik}(t+, 0+) [G_{0,kl}(0-, 0+)]^{-1} G_{0,lj}(0-, t^++).
\end{align}
Following Ref. \cite{Haldar}, it can be shown that the $n$-th R\'{e}nyi entropy is given by $S_A^{(n)}(t)=(1/(1-n))\mathrm{tr}\ln[(\mathbb{I}-\mathbb{C}(t))^n+\mathbb{C}^n(t)]$. As a result, by taking the limit $n\to 1$, we get the von Neumann entanglement entropy,
\begin{align}
S_A(t)=-\mathrm{tr}\left[(\mathbb{I}-\mathbb{C}(t))\ln(\mathbb{I}-\mathbb{C}(t))+\mathbb{C}(t)\ln\mathbb{C}(t)\right]. \label{eq:StateEE}
\end{align}
Thus we can obtain the growth of state entanglement entropies in terms of single-particle vacuum Green's functions (Appendix \ref{app:NIGF}).

%%%%%%%%%%%%%%%%%%%%%%%%%%%%%%%%%%%%%%%%%%%%%%%%%%%%%%%%%%%%%%%
\begin{figure}[H]
    \centering
    \includegraphics[width=0.75\linewidth]{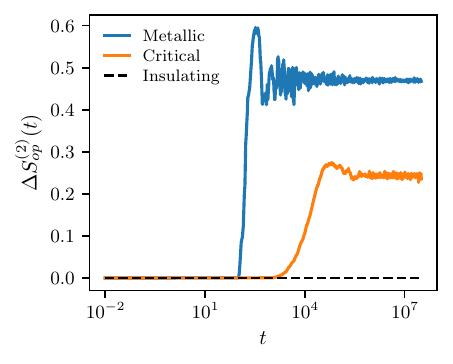}
    \caption{Growth of operator R\'{e}nyi entropy for quasiperiodic 1D Aubry-Andr\'{e} model in the ballistic metallic phase ($V=0.5$), at the sub-diffusive/nearly diffusive critical point ($V=1.0$), and in the localized insulating phase ($V>1$) for system size $L = 576$.}
    \label{fig:schematic2}
\end{figure}
%%%%%%%%%%%%%%%%%%%%%%%%%%%%%%%%%

%%%%%%%%%%%%%%%%%%%%%%%%%%%%%%%%%%%%%%%%%%%%%%%%%%%%%%%%%%%%%%%%%%
\section{Results for non-interacting quasiperiodic and random systems} \label{sec:Results}

In this section, we investigate three non-interacting models with well-characterized transport, ranging from ballistic to various sub-ballistic behaviors, like diffusive, super- and sub-diffusive, as well as Anderson localized. In particular, we study -- (1) 1D and (2) 2D quasiperiodic Aubry-Andr\'{e} (AA) models \cite{AA}, and (3) 2D Anderson model with random disorder. We compute the growth of $S_{n_lA}^{(2)}\equiv S_{op}^{(2)}$, $S_A^{(2)}\equiv S^{(2)}$ and $S_A=S^{(1)}$ by numerically evaluating Eqs.\eqref{eq:OperatorRenyi},\eqref{eq:StateRenyiEE},\eqref{eq:StateEE} using the correlation matrices in Eqs.\eqref{eq:OpCorrMatrix},\eqref{eq:StateCorrMatrix} for a given realization of the quasiperiodic or random potential. The entropies are further averaged over disorder realizations. We analyze how the growth and propagation of the operator R\'{e}nyi entropy are affected by ballistic and sub-ballistic transports, and localization transition in AA models. For example, $\Delta S_{op}^{(2)}(t)$ is shown in Fig.\ref{fig:schematic2} for 1D AA model across the localization transition from a ballistic metallic phase to an Anderson localized insulating phase, through a critical state with sub-diffusive or nearly-diffusive transport. As evident, $\Delta S_{op}^{(2)}(t)$ starts growing above a characteristic \emph{onset} time scale for the metallic and critical states and saturates to $O(1)$ value. The onset time and a related \emph{saturation} time (see later) depends on the distance of the initial local operator from the boundary of two subsystems (Fig.\ref{fig:schematic1}). We consider the bi-partition of the system into two halves $A$ and $B$, with subsystem length $L_A=L/2$. In 2D, we apply periodic boundary conditions in both directions. We place the local operator $\hat{n}_l$ at a site $l$ maximally distant from all the boundaries, with the distance linearly scaling with the system size $L$. We connect the system-size scaling of the saturation time with the known transport behaviors \cite{Sutradhar,Purkayastha} of AA models. In contrast, the operator R\'{e}nyi entropy does not grow in the localized phase in the absence of any transport as shown in Fig.~\ref{fig:schematic2}.

We also analyze early and intermediate time power-law growth of the half-subsystem state-entanglement entropies $S^{(2)}(t)$ and $S^{(1)}(t)$ with time, starting from a pure product state, namely the N\'{e}el state $|\psi(0)\rangle=|1010\cdots\rangle$, and connect to transport characteristics. After the temporal growth, the state entanglement entropies also eventually saturate to a steady-state value that scales with the volume of the subsystem in the metallic phase and at the critical point, whereas the steady state value is $O(1)$ in the localized phase. The scaling of a suitably defined saturation time scale with system size for the state entanglement entropies in the metallic phase and at the critical point of the AA models also confirms the relation with transport. 

We further study the operator R\'{e}nyi and state entanglement entropies in the finite-size diffusive regime $\ell\ll L\ll \xi$ of the 2D Anderson model with random disorder, where $\ell$ and $\xi$ are the mean free path and localization length, respectively.
%%%%%%%%%%%%%%%%%%%%%%%%%%%%%%%%%%%%%%%%%%%%%%%%%%%%%%%%%%%%%%%%%%

%%%%%%%%%%%%%%%%%%%%%%%%%%%%%%%%%%%%%%%%%%%%%%%%%%%%%%%%%%%%%%%%%%

\begin{figure*}[!htbp]
\centering
\includegraphics[width=0.9\textwidth]{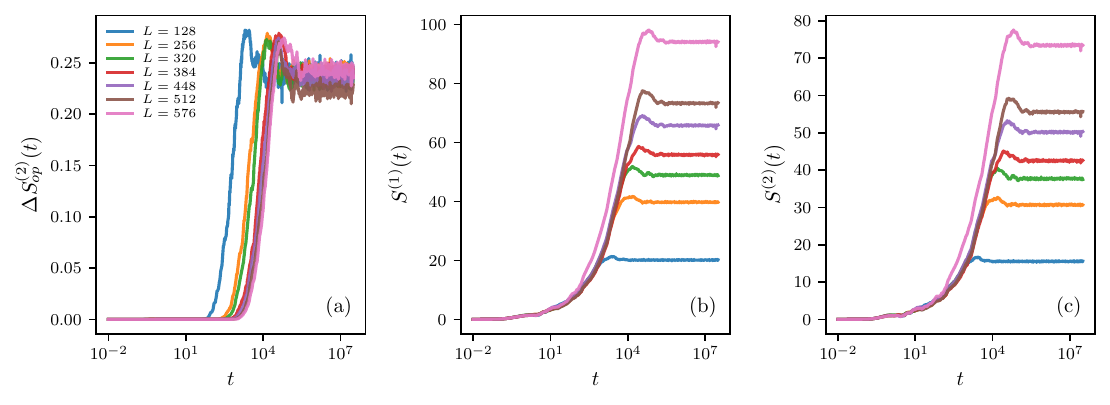}
\caption{Growth of (a) operator R\'{e}nyi entropy $S_{op}^{(2)}$, (b) von Neumann entanglement entropy $S^{(1)}$, and (c) second R\'{e}nyi entanglement entropy $S^{(2)}$ for 1D Aubry-Andr\'e Model at the critical point $V_C = 1$ for different system sizes $L=128-576$. The time evolutions of $S^{(1)}$ and $S^{(2)}$ have been computed starting from the N\'{e}el state.}
\label{fig:AA1D_V1}
\end{figure*}

\begin{figure*}[ht!]
\centering
\includegraphics[width=0.9\textwidth]{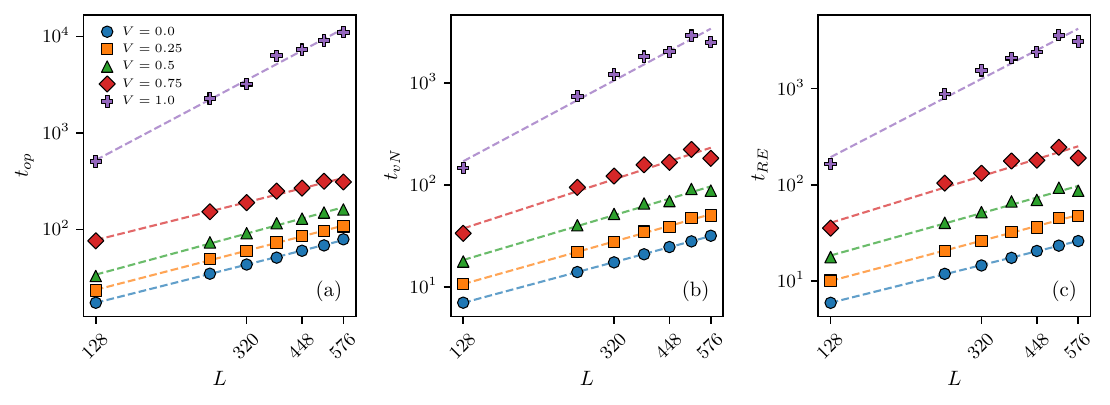}
\caption{Scaling of saturation times, (a) $t_{op}$, (b) $t_{vN}$, (c) $t_{RE}$, associated with the growth of the operator R\'{e}nyi entropy and state entanglement entropies, as a function of system size $L$ in the metallic phase and at the critical point, $V=0.0-1.0$ [legend in (a)] for the 1D Aubry-Andr\'e model.  The exponents for the power-law growth of the time scales with $L$ are shown in Table~\ref{tab:AA_1D_SAT}.}
\label{fig:AA1D_SAT}
\end{figure*}

\subsection{1D Aubry-Andr\'{e} Model}\label{sec:1DAA}
We first consider the 1D Aubry-Andr\'{e} model \cite{AA}, 
\begin{equation}
    \mathcal{H} =  \sum_r (c_r^\dagger c_{r + 1} + c_{r+1}^\dagger c_r) + 2V \sum_r \cos{\lf 2 \pi b r + \phi \rf} c_r^\dagger c_r,
\end{equation}
on a lattice with $L$ sites and periodic boundary condition, where $c_r^\dagger$ ($c_r$) are electron creation (annihilation) operator at site $r=1,\cdots,L$. Here we have set the hopping to 1, as the unit of energy/inverse time ($\hbar=1$). The on-site quasiperiodic potential is determined by the strength $V$ and an irrational number $b$, which we choose to be the golden ratio conjugate $b=(\sqrt{5}-1)/2$. The phase $\phi\in [0,2\pi)$, which induces a shift of the potential with respect to the lattice sites, is used to generate a statistical ensemble to perform \emph{disorder} averaging. The operator R\'{e}nyi and state entanglement entropies are averaged over 100 disorder realizations for all the system sizes.
The model is self-dual at $V=1$, through a duality mapping between real and momentum space \cite{AA,Deng2017}. The self-dual point also coincides with the metal-insulator transition between a ballistic metal for $V<1$ and localized insulator for $V>1$. 

The transport properties of the above non-interacting model, and its variants, \cite{Deng2017,Purkayastha} have been studied in several works \cite{Purkayastha2017,Sutradhar}, using various transport-related quantities, like spread of an initially localized wave packet with time \cite{Purkayastha} and system-size scaling of Thouless conductance \cite{Sutradhar} and Kubo conductivity for a \emph{closed or isolated system}. Similarly, for the \emph{open system} connected with leads at the two ends, system-size scaling of linear-response Landauer and Kubo conductances, and of steady-state current in the presence of a finite voltage bias \cite{Purkayastha} have been studied. These works unambiguously confirm the ballistic transport in the metallic phase and the absence of transport in the localized phase. However, the nature of transport at the critical point is more intricate and depends on the specific transport quantity studied, as well as whether the transport is studied for the ground state or infinite-temperature states. 

For the ground state, transport behavior can vary from sub-diffusive/nearly diffusive to strongly sub-diffusive depending on the single-particle state at the Fermi energy, as inferred from system-size dependence of the open-system Landauer and Kubo conductances \cite{Sutradhar}. The wave-packet spread and infinite-temperature Kubo conductivity in an isolated system indicate diffusive to weakly super-diffusive transport \cite{Purkayastha}. In contrast, open-system steady-state current at infinite temperature \cite{Purkayastha} and closed-system Thouless conductance \cite{Sutradhar} exhibit clear sub-diffusive scaling with system size. The distinct transport characteristics of various transport quantities at the critical point presumably arise due to the fractal nature of the energy spectrum \cite{Ostlund1983,Ostlund1984} and the resultant sensitivity of the single-particle wave function on the energy. Moreover, the multifractality of the single-particle wavefunctions at the critical point in the bulk and at the surfaces or the boundaries affect the closed and open-system properties differently \cite{Sutradhar}.
%%%%%%%%%%%%%%%%%%%%%%%%%%%%%%%%%%%%%%%%%%%%%%%%%%%%%%%%%%%%%%%%%%

%%%%%%%%%%%%%%%%%%%%%%%%%%%%%%%%%%%%%%%%%%%%%%%%%%%%%%%%%%%%%%%%%%
\begin{figure}[!htbp]
    \centering
    \includegraphics[width = 0.8\linewidth]{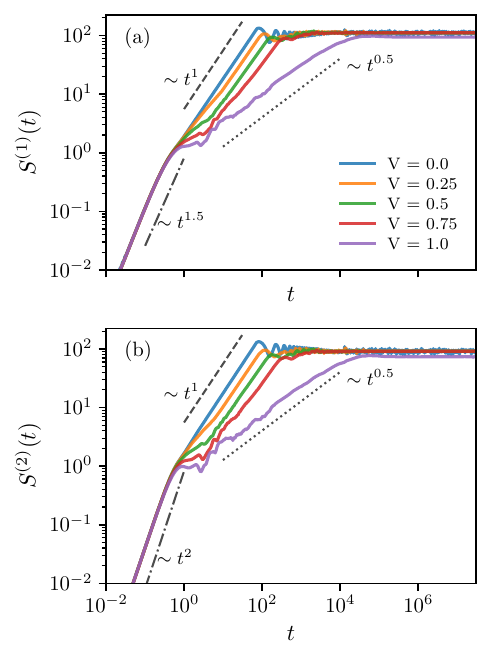}
    \caption{Time evolution of (a) von Neumann, and (b) second R\'{e}nyi entanglement entropy starting from an initial unentangled N\'{e}el state in the metallic phase and at the critical point ($V=0.0-1.0$) for the 1D Aubry-Andr\'e Model. The dashed lines are guides to the eye for the diffusive ($\sim \sqrt{t}$), ballistic ($\sim t$) and super-ballistic ($\sim t^2$) behaviors. %\textcolor{magenta}{SB: Make it a one-column figure with (a) top panel and (b) bottom panel.}
    }
    \label{fig:AA1D_L576}
\end{figure}

\subsubsection{Scaling of saturation time with system size}
To illustrate the dynamical behavior of the entropies, we present the time evolution of the operator R\'{e}nyi entropy, alongside the von Neumann and second R\'{e}nyi entanglement entropies for an initial N\'{e}el state, in Fig.~\ref{fig:AA1D_V1} at the critical point $V=V_c=1$. As shown in Fig.~\ref{fig:AA1D_V1}(a) for different system sizes $L=128-576$, the operator R\'{e}nyi entropy initially does not grow till a system-size-dependent onset time and then increases as $\sim \ln t$, eventually to saturate to an $O(1)$ value $S_{sat}$. Similar behaviour is found for other values of $V<1$ in the metallic phase (not shown). It can be shown \cite{Nozaki} that the growth of the operator R\'{e}nyi entropy is bounded by the local Hilbert space dimension, i.e., 2 in this case for spinless fermions, such that $\Delta S_{op}^{(2)}(t)\leq \ln 2$ (see Appendix \ref{app:Smaxbound}). 
% \textcolor{red}{For a clean tight-binding chain, we find that $S_{sat}=\ln 2$, as shown in Fig. \ref{fig:AA1D_Benchmark}(Appendix \ref{app:benchmark})}.
% \textcolor{cyan}{The long-time average can be lower than $\ln 2$}.
However, for the models with quasiperiodic and random potential considered here $S_{sat}<\ln2$ and $S_{sat}$ decreases with $V$. We define a saturation time $t_{op}(L)$ as the time required for $\Delta S_{op}^{(2)}(t)$ to reach half of the saturation value. We show in Fig.~\ref{fig:AA1D_SAT}(a) that $t_{op}\sim L^\alpha$, i.e., $t_{op}$ has power-law dependence on $L$ throughout the metallic phase ($V<1$) and at the critical point ($V=1$). The exponent $\alpha$, denoted as $\alpha_{op}$ in Table~\ref{tab:AA_1D_SAT}, is close to 1 throughout the metallic phase ($V<1$), implying a ballistic scaling between the saturation time $t_{op}$ and the distance ($\propto L$) of the operator $\hat{n}_l$ from the boundaries between $A$ and $B$ subsystems. At the critical point, we find nearly diffusive scaling, $t_{op}\sim L^2$, for the saturation time. Since we consider a system with periodic boundary condition, this diffusive scaling operator spread, as quantified by the operator R\'{e}nyi entropy $\Delta S_{op}^{(2)}(t)$, is consistent with diffusive spread of single-particle wave packet in an isolated 1D AA chain at $V=V_c=1$ \cite{Purkayastha}. The spread of initially localized (delta function) wave packet is characterized by the width of the wave packet as a function of time or the root mean square displacement $\sim \sqrt{t}$, similar to the scaling $\sqrt{t_{op}}\sim L$.

The temporal growth of the von Neumann and second R\'{e}nyi entanglement entropy starting from a N\'{e}el state is shown in Figs.~\ref{fig:AA1D_V1}(b),(c) at the critical point $V=1$. Similar behavior is also seen for other values of $V<1$ (not shown). As in the case of $t_{op}(L)$, saturation times, $t_{vN}$ and $t_{RE}$, are extracted by tracking the time at which $S^{(1)}(t)$ and $S^{(2)}(t)$ reach half of their eventual steady-state saturation values at long times. Unlike $\Delta S_{op}^{(2)}(t)$, whose steady-state value is $O(1)$ and depends on local Hilbert-space dimension, the steady-state values of the state entanglement entropies have a volume-law scaling with system size $L$ (not shown). As shown in Table~\ref{tab:AA_1D_SAT}, $t_{vN}$ ($\sim L^{\alpha_{vN}}$) and $t_{RE}$ ($\sim L^{\alpha_{RE}}$) exhibit ballistic scaling with $\alpha_{vN},\alpha_{RE}\simeq 1$ over most of the metallic phase, albeit, deviating towards weakly sub-ballistic scaling $\sim L^{1.2}$ close to the critical point, e.g., for $V=0.75$ (Table~\ref{tab:AA_1D_SAT}). At the critical point, $t_{vN}$ and $t_{RE}$ exhibit nearly diffusive scaling $\sim L^2$, identical to that of $t_{op}$.
\begin{table}[H]
\centering
\begin{tabular}{|c|c|c|c|}
\hline
V & \textbf{$\alpha_{op}$} & \textbf{$\alpha_{vN}$} & \textbf{$\alpha_{RE}$} \\
\hline
0     & 1.0  & 1.0  & 1.0  \\
0.25  & 1.0  & 1.1  & 1.1  \\
0.5   & 1.1  & 1.1  & 1.1  \\
0.75  & 1.0  & 1.2  & 1.2  \\
1     & 2.0  & 2.0  & 2.0  \\
\hline
\end{tabular}
\caption{The exponents $\alpha_{op},~\alpha_{vN},~\alpha{RE}$ for the power-law scaling of the saturation times $t_{op},~t_{vN},~t_{RE}$ in the metallic phase and at the critical point for the 1D Aubry-Andr\'e model.}
\label{tab:AA_1D_SAT}
\end{table}
%%%%%%%%%%%%%%%%%%%%%%%%%%%%%%%%%%%%%%%%%%%%%%%%%%%%%%%%%%%%%%%%%%

%%%%%%%%%%%%%%%%%%%%%%%%%%%%%%%%%%%%%%%%%%%%%%%%%%%%%%%%%%%%%%%%%%
\subsubsection{Temporal growth of state entanglement entropies}
As we have discussed above, the nature of transport can be diagnosed in the growth of the operator R\'{e}nyi entropy and state entanglement entropies through the scaling of a suitably defined saturation time with the (sub)system size. However, ballistic and sub-ballistic transport also give rise to characteristic power-law growth of state entanglement entropies as a function of time, in the early-to-intermediate time regime.
In {Fig.~\ref{fig:AA1D_L576}(a),(b), we present the temporal evolution of the von Neumann and R\'{e}nyi entanglement entropies, starting from the N\'{e}el state, for the largest system size considered, $L = 576$. Initially, over a short microscopic time interval $t\sim O(1)$, the von Neumann entanglement entropy grows super-ballistically as $S^{(1)}(t)\sim t^{1.5}$ for all values of $V\leq 1$, as shown in Fig.~\ref{fig:AA1D_L576}(a). After this interval, the von Neumann entanglement entropy grows ballistically as $\sim t$ in the ballistic metallic phase, whereas the growth is nearly diffusive $\sim \sqrt{t}$ at the critical point. For the second R\'{e}nyi entanglement entropy, the results reveal an initial super-ballistic growth ($\sim t^2$) over a brief microscopic timescale, independent of $V$. Subsequently, the growth of $S^{(2)}$, like that of the von Neumann entanglement entropy, reflects the underlying transport, exhibiting ballistic behavior ($\sim t$) in the metallic phase and sub-diffusive/nearly diffusive behavior ($\sim \sqrt{t}$) at the critical point. 

For clean non-interacting systems, the ballistic growth of entanglement entropy can be understood from the quasiparticle picture ~\cite{Calabrese2005,Calabrese2007}, where entangled quasiparticles are created locally due to quench of the initial product state under the Hamiltonian $\mathcal{H}$. One of the partners in a pair of quasiparticles, created initially in close proximity within a subsystem, ballistically propagates to the boundary of the subsystems and entangle a 
region of length $\propto t$ of the subsystems near their boundary. This leads to a ballistic growth of the entanglement entropy, albeit after a microscopic time scale over which entangled local quasiparticle pairs are created and their density is locally equilibrated to a value \cite{Rajabpour,Unanyan,Suh,Casini,Cotler}. 
 Thus, a super-ballistic growth of the entanglement entropy, e.g., $S^{(1)}\sim t^2$, over an early microscopic time of ``local equilibration", where the density of entangled quasiparticle pairs changes with time, can precede the ballistic growth at later times. The entanglement growth eventually saturates at a time scale $\sim L_A/2v$, where $v$ is the characteristic quasiparticle velocity. Even for the 
 quasiperiodic systems considered here we expect the quasiparticle picture of the ballistic growth of the entanglement entropy to be valid in the ballistic metallic phase for $V<1$. However, the initial microscopic super-ballistic growth [$S^{(1)}\sim t^{1.5}$, Fig.~\ref{fig:AA1D_L576}(a)] in the 1D Aubry-Andr\'{e} model for the von Neumann entanglement entropy is found to be different from $t^2$ growth. On the contrary, the second R\'{e}nyi entanglement entropy grows as $\sim t^2$ at early times, followed by a ballistic growth for $V<1$, as shown in Fig.~\ref{fig:AA1D_L576}(a).
 
% The behaviour of entanglement-growth at the critical point, where the transport is known to be sub-diffusive, agrees to the findings in \cite{Zhou_2020, Pollmann_2019, Huang_2020}, where it is argued that R\`enyi-entropies $S_{n>1}(t) \sim \sqrt{t}$ for non-integrable systems with diffusive transport. 

%%%%%%%%%%%%%%%%%%%%%%%%%%%%%%%%%%%%%%%%%%%%%%%%%%%%%%%%%%%%%%%%%%

\begin{figure*}[htbp!]
\centering
\includegraphics[width=0.9\textwidth]{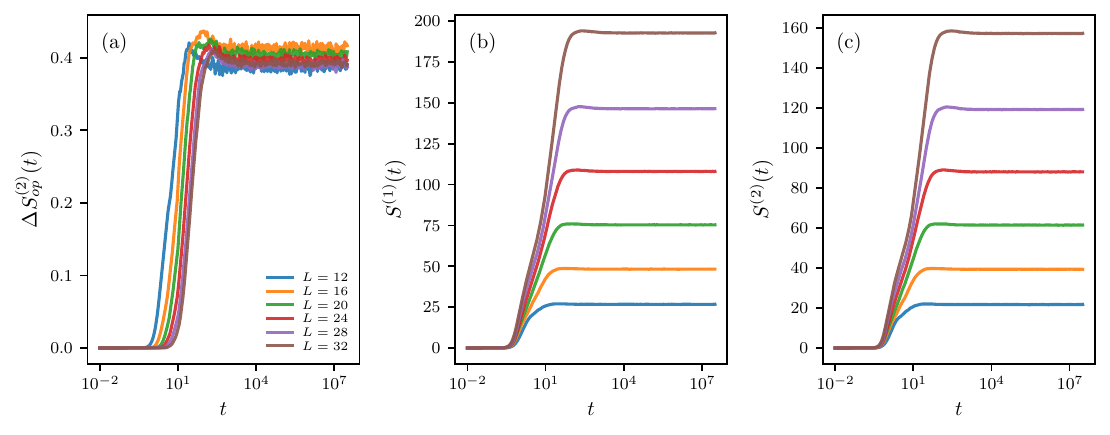}
\caption{Growth of (a) operator R\'{e}nyi entropy $S_{op}^{(2)}$, (b) von Neumann entanglement entropy $S^{(1)}$, and (c) second R\'{e}nyi entanglement entropy $S^{(2)}$ for the 2D Aubry-Andr\'e Model at the critical point $V_C = 1$ for different system sizes $L=12-32$ with no next-nearest-neighbor hopping ($t'=0$). The time evolutions of $S^{(1)}$ and $S^{(2)}$ have been computed starting from the N\'{e}el state.}
\label{fig:AA2D_V1}
\end{figure*}

\begin{figure*}[htbp!]
\centering
\includegraphics[width=0.9\textwidth]{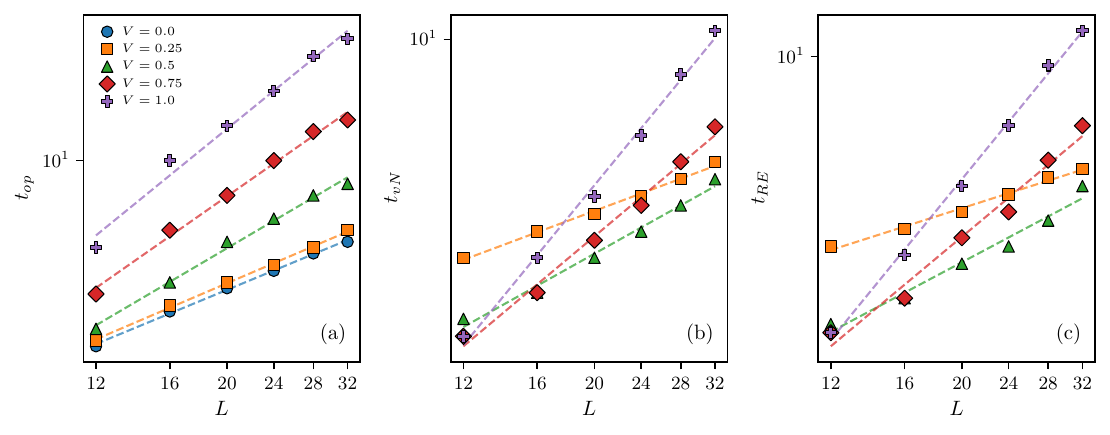}
\caption{Scaling of saturation times, (a) $t_{op}$, (b) $t_{vN}$, (c) $t_{RE}$, associated with the growth of the operator R\'{e}nyi entropy and state entanglement entropies, as a function of system size $L$ in the metallic phase and at the critical point, $V=0.0-1.0$ for the 2D Aubry-Andr\'e model with no next-nearest-neighbor hopping ($t'=0$). The exponents for the power-law growth of the time scales with $L$ are shown in Table~\ref{tab:AA_2D_SAT}.}
\label{fig:AA2D_SAT}
\end{figure*}
%%%%%%%%%%%%%%%%%

\subsection{2D Aubry-Andr\'e Model}
The preceding sections show that ballistic transport in the metallic phase and nearly diffusive transport at the critical point of the 1D Aubry-Andr\'{e} model are very clearly captured through the growth of operator R\'{e}nyi and state entanglement entropies. In this section, we investigate the same quantities in the 2D generalization \cite{Devakul} of the 1D Aubry-Andr\'{e} model, namely the non-interacting 2D Aubry-Andr\'{e} model with the Hamiltonian, 
\begin{align}
\mathcal{H} &= \sum_{\mathbf{r}, \mu} \left( e^{i \phi_\mu} c^\dagger_{\mathbf{r} + \hat{{\mu}}} c_{\mathbf{r}} + \mathrm{h.c.} \right) + \sum_{\mathbf{r}} \epsilon_{\mathbf{r}} c^\dagger_{\mathbf{r}} c_{\mathbf{r}} \label{eq:modelHamiltonian} \\
\epsilon_{\mathbf{r}} &= 2V \sum_{\mu} \cos \left( 2\pi \sum_{\nu} B_{\mu \nu} r_\nu + \phi_\mu \right) \label{eq:potential}
\end{align}
where $c_{\mathbf{r}}$ represents the fermion annihilation operator at site $\mathbf{r}$ on a 2D square lattice with $L\times L$ sites, and $\mu,\nu = 1,2$ index the Cartesian directions. We define the matrix $\mathbb{B} = b \mathbb{R}$, with $b = (\sqrt{5}-1)/2$ and $\mathbb{R}$ an orthonormal matrix, as specified in Ref.~\onlinecite{Devakul}. The matrix $\mathbb{R}$ is given by
\begin{align}
\mathbb{R} &= \begin{bmatrix}
c & -s \\
s & c
\end{bmatrix},
% \mathbb{R} &= \begin{bmatrix}
% c^2 + s^3 & cs & cs^2 - cs \\
% cs & -s & c^2 \\
% cs^2 - cs & c^2 & c^2 s + s^2
% \end{bmatrix}, \quad d = 3
\end{align}
where $c = \cos\theta$ and $s = \sin\theta$. For all calculations, we adopt $\theta = \pi/7$. The phases $\phi_\mu$ are again used to generate a statistical ensemble for disorder averaging over 100 disorder realizations to obtain disorder-averaged operator R\'{e}nyi and state entanglement entropies. Like the 1D Aubry-Andr\'{e} model, the above model is self-dual at $V=V_c=1$, which coincides with the metal-insulator critical point. In Ref.~\onlinecite{Sutradhar}, transport properties of the above self-dual 2D Aubry-Andr\'e model were studied across the localization transition via open-system Kubo conductance. The latter indicated ballistic to super-diffusive crossover in the metallic phase with increasing $V$, and sub-diffusive transport at the critical point. However, these transport regimes only ensue for system size $L\gg\ell(V)$, where $\ell(V)$ is a microscopic length that can vary from 50 to 500 with increasing $V$ towards the critical point \cite{Sutradhar}. For $L\ll \ell$, the Kubo conductance exhibits diffusive scaling with system size. For reasons discussed later, we also study the growth of operator R\'{e}nyi and state entanglement entropies by adding a next-nearest-neighbor hopping $t'$ to the Hamiltonian in Eq.\eqref{eq:modelHamiltonian}. The next-nearest-neighbor hopping breaks the self-duality of the 2D Aubry-Andr\'{e} model.

%%%%%%%%%%%%%%%%%%%%%%%%%%%%%%%%%%%%%%%%%%%%%%%%%%%%%%%%%%%%%%%%%%

%%%%%%%%%%%%%%%%%%%%%%%%%%%%%%%%%%%%%%%%%%%%%%%%%%%%%%%%%%%%%%%%%%
\subsubsection{Scaling of saturation time}
To check whether the growth of the operator R\'{e}nyi and state entanglement entropies is able to capture the transport behaviors in the self-dual 2D Aubry-Andr\'{e} model for $t'=0$, particularly the ballistic to super-diffusive crossover for $V<1$ and sub-diffusive transport at $V=1$, we analyze the temporal dynamics of the entropies in the metallic phase ($V<1$) up to the critical point ($V=1$). In Fig.~\ref{fig:AA2D_V1}(a),(b),(c), we plot $\Delta S_{op}^{(2)}(t)$, $S^{(1)}(t)$ and $S^{(2)}(t)$ at the critical point $V = V_c = 1$ for various system sizes $L=12-32$. Other values of  $V<1$ also produce similar plots (not shown). The operator R\'{e}nyi entropy [Fig.~\ref{fig:AA2D_V1}(a)] shows similar behavior as in the 1D case; a system-size-dependent time for the onset of growth, followed by a $\sim \ln t$ growth, and eventually a saturation at a steady-state $O(1)$ value. 

Contrary to the 1D case [Fig.~\ref{fig:AA1D_V1}(b),(c)], the half-system von Neumann and second R\'{e}nyi entanglement entropies for an initial N\'{e}el state start growing with a power law in time (see later) only after a system-size-independent onset time. The entanglement entropies eventually saturate to a value scaling as $\sim L^2$, i.e., as a volume-law with subsystem size. 
\begin{figure}[!htbp]
    \centering 
    \includegraphics[width = 0.8\linewidth]{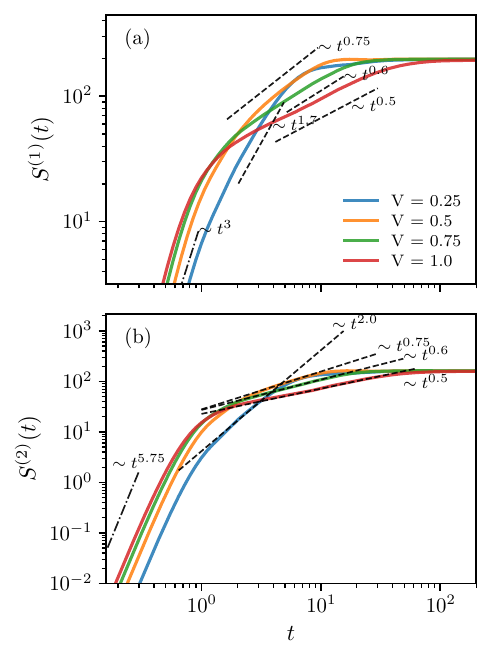}
    \caption{Time evolution of (a) von Neumann, and (b) second R\'{e}nyi entanglement entropy for an initial N\'{e}el state for the metallic phase ($V<1$) and at the critical point ($V=1$) for $L=32$ in 2D Aubry-Andr\'e Model. The dashed lines are guides to the eye for the diffusive, super-diffusive, ballistic and super-ballistic behaviors. %\textcolor{magenta}{SB: Add the figure for von Neumann entanglement entropy. Make it a one-column figure with (a) von Neumann in top panel and (b) Renyi in the bottom panel.}
    }
    \label{fig:AA2D_L32}
\end{figure}
We again extract the saturation times, $t_{op}$, $t_{vN}$ and $t_{RE}$, as in the 1D Aubry-Andr\'{e} model (Sec.\ref{sec:1DAA}), and plot them as a function of $L$ in Fig.~\ref{fig:AA2D_SAT}(a),(b),(c). The saturation times exhibit power-law dependence ($\sim L^\alpha$) on $L$. The exponents are listed in Table~\ref{tab:AA_2D_SAT}. As evident, the critical point $V=1$ exhibits diffusive growth with $\alpha_{op},\alpha_{vN},\alpha_{RE}\simeq 2$, instead of asymptotic sub-diffusive transport found in Ref.~\onlinecite{Sutradhar} for $L\gg \ell$. This is consistent with the system sizes $L<\ell$ studied in this work. The exponent $\alpha_{op}$, lying between 1 and 2, shows the expected ballistic to super-diffusive crossover \cite{Sutradhar} in the metallic phase ($V<1$). 

The situation is not clear for the scaling of $t_{vN}$ and $t_{RE}$ with $L$ in the metallic phase. First of all, we find that state entanglement entropies do not grow at all for $V=0$ up to times studied here. As a result, we could not extract the exponents $\alpha_{vN}$ and $\alpha_{RE}$ at $V=0$. We find that adding a next-nearest-neighbor hopping, e.g., $t'=1/3$ on the 2D square lattice changes this situation and leads to a growth of state entanglement entropies with  a ballistic scaling $t_{vN},t_{RE}\sim L$ for $V=0$ (see Appendix~\ref{app:AA2D_NNN}). As shown in Table~\ref{tab:AA_2D_SAT}, in the absence of next-nearest-neighbor hopping, $t_{vN}$ and $t_{RE}$ exhibit anomalous super-ballistic scaling with $L$ for $V=0.25$ with $\alpha_{vN},\alpha_{RE}\simeq 0.5-0.6$, whereas $V=0.5$ and $V=0.75$ show close to ballistic scaling $\alpha_{vN},\alpha_{RE}\simeq 0.9$ and super-diffusive scaling $\alpha_{vN},\alpha_{RE}\simeq 1.4$, respectively. Again, we find that the addition of a next-nearest-neighbor hopping $t'=1/3$ changes the exponents to nearly ballistic ones, $\alpha_{vN},\alpha_{RE}\simeq 0.8-0.9$ for $V=0.25$ (see Fig.~\ref{fig:AA2D_NNN_SAT}, Table~\ref{tab:AA2D_NNN_SAT} in Appendix~\ref{app:AA2D_NNN}). The reason behind the lack of growth for operator R\'{e}nyi entropy in the nearest neighbor model at $V=0$, i.e., for the nearest-neighbor tight-binding model, and the \emph{anomalous} super-ballistic growth for small values of $V$, like $V=0.25$ is not clear. The anomalous behavior may have origin in the van-Hove singularity in the single-particle density of states, and the resultant dominant contributions of quasiparticle with vanishing velocities, in the clean limit ($V=0$) or the proximity to it. The super-ballistic behavior is also seen in the temporal growth of state entanglement entropies for small values of $V$ in the nearest-neighbor 2D Aubry-Andr\'{e} model, as discussed in the next section.

% The growth phase is $\sim \ln t$ \cite{Alba_2020, Prosen_2009}. It is seen that for $V=0$ the state-entanglements does not grow for timescales we consider numerically. For this reason, we don't consider this value of $V$ for extracting system-size dependence of saturation-time $t_S$. We tabulate the exponents $\alpha(V)$ in Table ~ \ref{tab:AA_2D_SAT}. The ballistic to super-diffusive crossover \cite{Sutradhar_2018, Abhishek_2018} in the metallic phase is well-captured in case of operator entanglement.

\begin{table}[H]
\centering
\begin{tabular}{|c|c|c|c|}
\hline
V & \textbf{$\alpha_{op}$} & \textbf{$\alpha_{vN}$} & \textbf{$\alpha_{RE}$} \\
\hline
0     & 1.0  & $\times$  & $\times$  \\
0.25  & 1.0  & 0.6  & 0.5  \\
0.5   & 1.4  & 0.9  & 0.9  \\
0.75  & 1.7  & 1.4  & 1.4  \\
1     & 2.0  & 2.0  & 2.0  \\
\hline
\end{tabular}
\caption{The exponents $\alpha_{op},~\alpha_{vN},~\alpha{RE}$ for the power-law scaling of the saturation times $t_{op},~t_{vN},~t_{RE}$ in the metallic phase and at the critical point for the 2D Aubry-Andr\'e model with no next-nearest-neighbor hopping ($t'=0$).}
\label{tab:AA_2D_SAT}
\end{table}
%%%%%%%%%%%%%%%%%%%%%%%%%%%%%%%%%%%%%%%%%%%%%%%%%%%%%%%%%%%%%%%%%%

%%%%%%%%%%%%%%%%%%%%%%%%%%%%%%%%%%%%%%%%%%%%%%%%%%%%%%%%%%%%%%%%%%
\subsubsection{Temporal growth of state entanglement entropies} \label{sec:2DAA_TemporalGrowth}
In contrast to the scaling of the saturation times for state entanglement entropies with system size, the temporal power-law growth of the state entanglement entropies provides a clearer connection to transport. In Fig.~\ref{fig:AA2D_L32}(a),(b), we present the temporal evolution of the von Neumann and the second R\'enyi entanglement entropies for an initial N\'{e}el state for the largest system size considered, $L = 32$. The results reveal early-time super-ballistic growth, $S^{(1)}\sim t^{3}$ and $S^{(2)}\sim t^{5.8}$, over a brief microscopic timescale, independent of $V$. After this, an intermediate time power-law growth with $t$ ensues for both the entropies, varying from diffusive ($\sim \sqrt{t}$) at the critical point to super-diffusive ($\sim t^\alpha$, $0.5<\alpha<1$) for most of the metallic regime, except for small $V$ (0.25), where the growth is super-ballistic, $S^{(1)}\sim t^{1.7}$ and $S^{(2)}\sim t^2$, as shown in Fig.~\ref{fig:AA2D_L32}(a),(b). Again, adding a next-nearest-neighbor hopping $t'=1/3$ for $V=0.25$ changes the behaviors to more expected ones, $S^{(1)}\sim t^{1.5}$ and $S^{(2)}\sim t^2$ at early microscopic time, followed by ballistic growth, as shown in Fig.~\ref{fig:AA2D_NNN_L32}, Appendix \ref{app:AA2D_NNN}.

Thus, generically, the state entanglement entropies capture the intermediate-time ballistic to super-diffusive crossover in the metallic phase, as expected from transport \cite{Sutradhar}. The early microscopic-time super-ballistic growth and intermediate-time super-diffusive growth of the state entanglement entropies in the metallic phase of the 2D Aubry-Andr\'{e} model can be rationalized by generalizing the quasiparticle picture \cite{Calabrese2005,Rajabpour,Unanyan}, discussed in the preceding section for the 1D Aubry-Andr\'{e} model, by taking into account the super-diffusive spread of the quasiparticle wave packet.

\begin{figure*}[htbp!]
\centering
\includegraphics[width=0.9\textwidth]{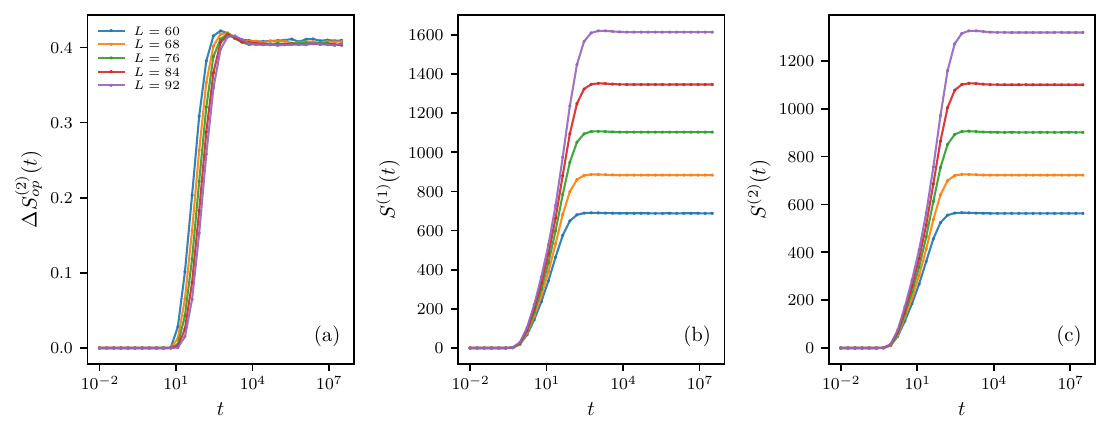}
\caption{Growth of (a) operator R\'{e}nyi entropy $S_{op}^{(2)}$, (b) von Neumann entanglement entropy $S^{(1)}$, and (c) second R\'{e}nyi entanglement entropy $S^{(2)}$ for the 2D Anderson model with $W=2.0$ for different system sizes $L=60-92$. The time evolutions of $S^{(1)}$ and $S^{(2)}$ have been computed for an initial N\'{e}el state.}
\label{fig:ANDERSON2D_V2}
\end{figure*}

\begin{figure}[htbp!]
\centering
\includegraphics[width = 0.8\linewidth]{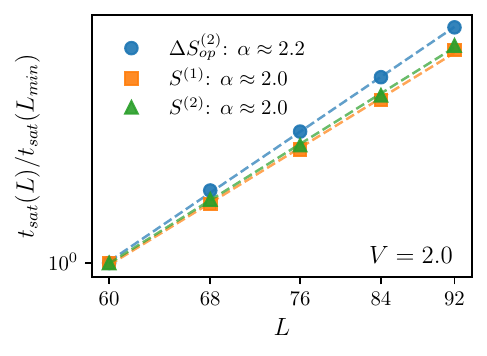}
\caption{Scaling of saturation times, (a) $t_{op}$, (b) $t_{vN}$, (c) $t_{RE}$, associated with the growth of the operator R\'{e}nyi entropy and state entanglement entropies, as a function of system size $L$ at $W=2.0$ for the 2D Anderson model. The exponents for the power-law growth of the time scales with $L$ are shown in Table~\ref{tab:ANDERSON2D_SAT}. Saturation times $t_{\text{sat}}$ are rescaled by $t_{\text{sat}}(L_{\text{min}})$ to visualize the relative scaling behavior. Dashed lines indicate fits to the power law $t_{\text{sat}} \propto L^\alpha$. 
%\textcolor{magenta}{SB: Make this a one-column figure with $t_{op},t_{vN},t_{RE}$ plotted in the same plot. Scale the time scale accordingly to put in the same plot. Legend can indicate the time scale and if any scale factor is used.}
}
\label{fig:ANDERSON2D_SAT}
\end{figure}
%%%%%%%%%%%%%%%%%%%%%%%%%%%%%%%%%%%%%%%%%%%%%%%%%%%%%%%%%%%%%%%%%%

%%%%%%%%%%%%%%%%%%%%%%%%%%%%%%%%%%%%%%%%%%%%%%%%%%%%%%%%%%%%%%%%%%
\subsection{2D Anderson Model}\label{sec.2DAnderson}
To verify whether the operator R\'{e}nyi entropy and the state entanglement entropies capture the transport behavior beyond the quasiperiodic models, we consider the 2D Anderson model with random disorder, as described by the Hamiltonian,
\begin{align}
\mathcal{H} &= \sum_{\mathbf{r}, \mu} \left( c^\dagger_{\mathbf{r} + \hat{{\mu}}} c_{\mathbf{r}} + \mathrm{h.c.} \right) + \sum_{\mathbf{r}} \epsilon_{\mathbf{r}} c^\dagger_{\mathbf{r}} c_{\mathbf{r}}, \label{eq:modelHamiltonian} 
\end{align}
where $\epsilon_{\mathbf{r}} \in [-W, W]$ is a uniform random disorder potential with strength $W$. Strictly, the above random 2D system in the orthogonal symmetry class \cite{Abrahams1979,Evers2008} is always localized for any finite disorder strength due to weak localization. However, the localization length $\xi$ is exponentially large in the mean free path $\ell$ in 2D. As a result, we choose the disorder strength $W$ and system sizes $L$ such that the finite-size system of length $L$ in the diffusive transport regime, $\ell \ll L \ll \xi$. We take $W=2$, which leads to $\ell\sim O(1)$ and $\xi\approx 500$  [\onlinecite{MacKinnon}] for the 2D Anderson model on the square lattice with unit hopping and unit lattice spacing. Here we choose the finite-size diffusive regime in the weakly localized insulating phase of the 2D Anderson model to access larger linear size $L$ compared the 3D Anderson model, which has a proper diffusive metallic phase in the thermodynamic limit.

The results for $\Delta S_{op}^{(2)}(t)$, $S^{(1)}(t)$ and $S^{(2)}(t)$, averaged over 100 disorder realizations, are shown in Fig.~\ref{fig:ANDERSON2D_V2}(a),(b),(c) for disorder strength $W=2$. The growth of the entropies is very similar to that for the 2D Aubry-Andr\'{e} model discussed in the preceding section. The system size scaling of the saturation times extracted from the temporal growth of the entropies is shown in Fig.~\ref{fig:ANDERSON2D_SAT}. As listed in the Table~\ref{tab:ANDERSON2D_SAT}, the exponents $\alpha_{vN}, \alpha_{RE}\simeq 2$, implying diffusive scaling. The exponent $\alpha_{op}\simeq 2.2$ obtained from the operator R\'{e}nyi entropy turns out to be slightly sub-diffusive. A diffusive temporal power-law growth ($\sim \sqrt{t}$) regime can also be detected in $S^{(2)}(t)$, as shown in Fig. \ref{fig:ANDERSON2D_L92}. 

% We see here also the operator entanglement has a logarithmic early-time growth. As shown in Fig.~\ref{fig:ANDERSON2D_SAT}, after an initial growth of brief microscopic timescale, the entanglement grows diffusively $\sim t^{1/2}$ for almost two decades. The exponents $\alpha(W)$ for saturation-time scaling are recorded in Table ~\ref{tab:ANDERSON2D_SAT}.

\begin{table}[H]
\centering
\begin{tabular}{|c|c|c|c|}
\hline
W & \textbf{$\alpha_{op}$} & \textbf{$\alpha_{vN}$} & \textbf{$\alpha_{RE}$} \\
\hline
2.00     & 2.2  & $2.0$  & $2.0$  \\

\hline
\end{tabular}
\caption{$\alpha$ for different entanglement measures with disorder strength $W$ for the 2D Anderson model.}
\label{tab:ANDERSON2D_SAT}
\end{table}

\begin{figure}[htbp!]
    \centering
    \includegraphics[width = 0.8\linewidth]{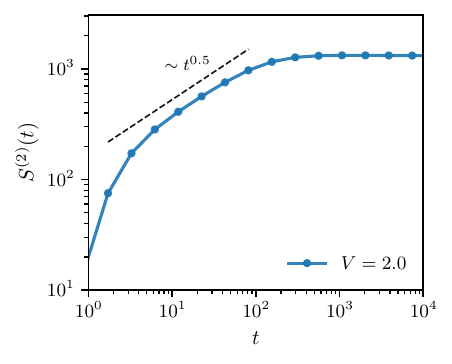}
    \caption{Time evolution of the second R\'{e}nyi entanglement entropy for an initial N\'{e}el state for $W=2$ in the 2D Anderson model for $L=92$. The dashed line is guide to the eye for the diffusive behavior. State entanglement grows diffusively for the disorder and system size chosen. %\textcolor{magenta}{SB: Modify the figure, i.e., x and y ranges, such that intermediate diffusive regime is amplified and more clear. The figure is dominated by the early-time regime, which is of much less importance here.}
    }
    \label{fig:ANDERSON2D_L92}
\end{figure}
%%%%%%%%%%%%%%%%%%%%%%%%%%%%%%%%%%%%%%%%%%%%%%%%%%%%%%%%%%%%%%%%%%

%%%%%%%%%%%%%%%%%%%%%%%%%%%%%%%%%%%%%%%%%%%%%%%%%%%%%%%%%%%%%%%%%%
\subsection{Temporal evolution of Schmidt values}

As discussed in the introduction, the diffusive and ballistic growths of state R\'{e}nyi and von Neumann entanglement entropies, respectively, for interacting diffusive systems, and ballistic growth of all the entropies in integrable systems and systems without conservation laws can be rationalized in terms of Schmidt values of the subsystem reduced density matrix \cite{PollMann, Ludwig,Rakovszky2019}.
Here we investigate temporal behavior of the half-subsystem many-body Schmidt values for the time evolution of the initial N\'{e}el state-entanglement in the 1D Aubry-Andr'{e} model $V = 0$ and at the critical point $V = 1$ and also for the 2D Aubry-Andr'{e} model $V = 0.75$. For non-interacting systems, and an initial Gaussian state like the N\'{e}el state, the reduced density matrix of a subsystem can be written as 
\begin{align}
\rho_A(t)=\frac{1}{\mathcal{Z}_A(t)}e^{-\sum_{i,j\in A}H_{ij}(t)c_i^\dagger c_j},  
\end{align}
where $\mathcal{Z}_A(t)=\mathrm{Tr}[\exp(-\sum_{i,j\in A}H_{ij}(t)c_i^\dagger c_j)]$ for each disorder realization. The matrix $\mathbb{H}(t)$ is obtained from correlation matrix $\mathbb{C}(t)$ of Eq.\eqref{eq:StateCorrMatrix} using $\mathbb{C}(t)=[1+\exp(\mathbb{H}(t))]^{-1}$. To this end, we first diagonalize $\mathbb{C}(t)$ for the half subsystem to obtain its eigenvalues $c_m(t)$ ($m=1,2,\cdots,L/2$) and consequently the eigenvalues $h_m(t)$ of $\mathbb{H}(t)$. The many-body Schmidt values are obtained by different combinations of the occupations $i\equiv \{n_m\}$ ($n_m=0,1$) of the states $m=1,\cdots,L/2$, i.e.,
\begin{align}
\Lambda_i=\frac{e^{-\sum_m h_m n_m}}{\prod_m(1+e^{-h_m})}
\end{align}

We numerically compute the ten largest many-body Schmidt values, $\Lambda_i(t)$ as function of time with $i=1,2,\ldots,10$, ordered as 
$\Lambda_1 > \Lambda_2 > \cdots > \Lambda_{10}$ for each disorder realization. 
We then obtain their typical values through geometric mean over $N_r = 100$ disorder realizations, defined as
\begin{equation*}
\Lambda_i^{\mathrm{typ}}(t) 
= \exp\!\left[
\frac{1}{N_r}
\sum_{r=1}^{N_r} 
\ln \Lambda_i^{(r)}(t)
\right].
\end{equation*}
Here $\Lambda_i^{(r)}(t)$ denotes the $i$-th Schmidt value for the $r$-th disorder realization.

\begin{figure*}[htbp!]
    \centering
    \includegraphics[width =0.98  \linewidth]{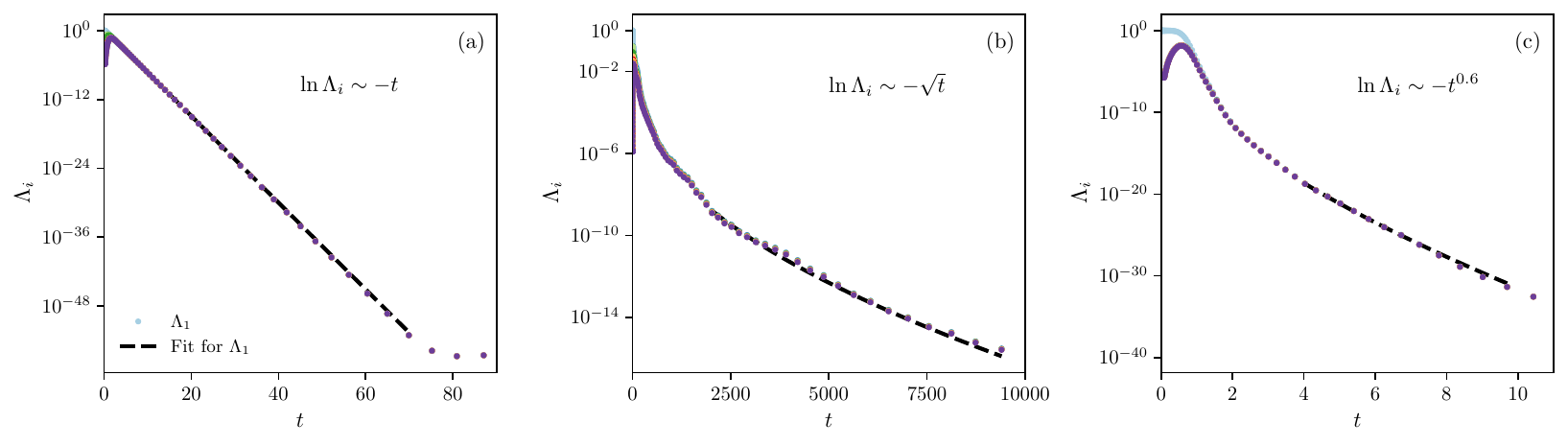}
    \caption{The time evolution of the ten largest Schmidt values $\Lambda_i(t)$ ($i=1,2,\cdots,10$) for initial N\'{e}el state in the 1D Aubry-Andr\'{e} model for (a) $V=0$ and (b) $V=1$ and 2D Aubry-Andr\'{e} model for (c) $V=0.75$. The data is fitted (dashed line) in the intermediate-to-long-time window with $\ln\Lambda_i\sim -t$ in (a) and with $\ln\Lambda_i\sim -\sqrt{t}$ in (b) and with $\ln\Lambda_i\sim -{t^{0.6}}$ in (c).}
    \label{fig:Schmidt_Decay}
\end{figure*}

% \begin{figure*}[htbp!]
%     \centering
%     \includegraphics[width = 0.8 \linewidth]{Schmidt_Fit_Parameters_Comparison.pdf}
%     \caption{\textcolor{magenta}{SB: Relevance of these figures is not clear. What insight do we get from these figures? Nothing is written either in the main text or in the caption, This can moved to appendix with some discussions in the text and caption so that their relevance is clear.}}
%     \label{fig:Schmidt_Fit_Params}
% \end{figure*}

The ten largest disorder-averaged Schmidt values are plotted in Fig.~\ref{fig:Schmidt_Decay} for the 1D Aubry-Andr\'{e} model, (a) in the metallic phase at $V=0$ with ballistic transport, and ballistic growth of operator R\'{e}nyi entropy and state entanglement [Table \ref{tab:AA_1D_SAT} and Fig.~\ref{fig:AA1D_L576}], and (b) at the critical point $V=1$ with nearly diffusive transport, and diffusive operator and entanglement growth [Table \ref{tab:AA_1D_SAT} and Fig.~\ref{fig:AA1D_L576}]. We also plot the Schmidt values for the 2D Aubry-Andr\'{e} model in Fig.~\ref{fig:Schmidt_Decay}(c) in the metallic phase at $V=0.75$ with super-diffusive transport and entropy growth [Table \ref{tab:AA_2D_SAT} and Fig.~\ref{fig:AA2D_L32}]. As shown in Figs.~\ref{fig:Schmidt_Decay}(a),(b), for the 1D Aubry-Andr\'{e} model, after some early-time intervals, all the Schmidt values decay in a ballistic fashion, i.e., exponentially as $\Lambda_i(t)=c_i e^{-d_i t}$ for $V=0$, and diffusively, i.e., slower than exponentially as $\Lambda_i(t)=c_i e^{-d_i \sqrt{t}}$ for $V=1$. In the 2D Aubry-Andr\'{e} model, as shown in Fig.~\ref{fig:Schmidt_Decay}(c), $\Lambda_i(t)=c_i e^{-d_i t^{0.6}}$ for $V=0.75$ in, indicating super-diffusive decay. We also show  in Fig.~\ref{fig:AA2D_Schmidt_V0p25} of Appendix~\ref{app:Schmidt} that the Schmidt values decay super-ballistically as $\Lambda_i(t)=c_i e^{-d_i t^{2}}$ for $V=0.25$, in the 2D Aubry-Andr\'{e} model, consistent with the anomalous super-ballistic operator and entanglement growth [Table \ref{tab:AA_2D_SAT} and Fig.~\ref{fig:AA2D_L32}]. The variation of the parameters $c_i$ and $d_i$ with Schmidt index $i$ are shown in Fig.~\ref{fig:AA1D_Schmidt_Fit}, Appendix~\ref{app:Schmidt}.

% \textcolor{red}{These shows qualitative difference between the contribution of Schmidt values in ballistic and diffusive regime}. (\textcolor{magenta}{SB: What qualitative difference? Need to explain better and more explicitly.}) 

Our results here in the non-interacting 1D Aubry-Andr\'{e}
 model demonstrate qualitatively distinct behavior of the Schmidt values for ballistic, super-diffusive and sub-diffusive/nearly diffusive transport. The time evolutions of the different Schmidt values at the nearly diffusive critical point are also unlike those for the interacting diffusive system \cite{PollMann,Ludwig,Rakovszky2019} where only the maximum Schmidt value decays diffusively, while all the other Schmidt values decay ballistically.

%%%%%%%%%%%%%%%%%%%%%%%%%%%%%%%%%%%%%%%%%%%%%%%%%%%%%%%%%%%%%%%

%%%%%%%%%%%%%%%%%%%%%%%%%%%%%%%%%%%%%%%%%%%%%%%%%%%%%%%%%%%%%%%
\section{Conclusions and Discussion} \label{sec:Conclusion}

In this work, we have proposed a subsystem operator R\'{e}nyi entropy as a state-independent measure of operator growth. By employing coherent-state path-integral representations, e.g., that of reduced subsystem operators in terms of fermionic displacement operators \cite{Haldar} and occupation basis product states \cite{Perugu}, we have developed a versatile Schwinger-Keldysh field-theoretic framework to compute both the operator R\'{e}nyi entropy, and state entanglement entropies for arbitrary initial pure product states in fermionic systems. We have further applied the method to obtain expressions for these operator and state entanglement entropies for non-interacting systems in terms of usual Keldysh Green's functions.

% By employing a coherent-state path-integral representation and leveraging the fermionic displacement operator formalism introduced in Ref.~\cite{Haldar_2020}, we have addressed the computational challenges associated with connecting multiple replicas in path-integral calculations. This approach provides a versatile and efficient tool for analyzing entanglement dynamics in lattice-based many-body systems, surpassing the limitations of previous conformal field theory methods.

Our numerical results for 1D and 2D quasiperiodic Aubry-Andr\'{e} models and the 2D Anderson model, establish a robust connection between entanglement and operator dynamics, and underlying transport properties. Specifically, we observe that operator and state entanglement entropies exhibit distinct temporal behaviors reflecting ballistic, sub-diffusive/nearly diffusive and super-diffusive quantum information propagation in unison with transport. In all the models, the subsystem operator R\'{e}nyi entropy displays a logarithmic growth phase followed by saturation, with the saturation timescale scaling as $t_S \sim L^\alpha$, with the exponent $\alpha\simeq 1$ in the ballistic metallic phase and $\alpha \simeq 2$ at the sub-diffusive/nearly diffusive critical point $V=1$ in the 1D Aubry-Andr\'{e} model. Similarly, in the 2D Aubry-Andr\'{e} model, the operator and entanglement growth capture a crossover from ballistic to super-diffusive behavior, while the 2D Anderson model exhibits diffusive growth for system sizes shorter than the localization length. Moreover, we show that all many-body Schmidt values of the reduced density matrix inherit the characteristic transport behavior by exhibiting corresponding ballistic or sub-ballistic temporal decay, including diffusive and anomalous diffusive behaviors. This is unlike diffusive interacting systems \cite{Ludwig, PollMann,Rakovszky2019} which exhibit dichotomy between maximum and typical Schmidt values, i.e., diffusive decay of the maximum Schmidt value as opposed to the ballistic decay of the typical eigenvalues.

The established link between entanglement and operator growth and transport properties has profound implications for understanding quantum information propagation in condensed matter systems. Our Schwinger-Keldysh framework not only unifies the computation of operator and state entanglement entropies but also elucidates their complementary roles in probing quantum correlations. Crucially, the path-integral representation developed here paves the way for extending the formalism to interacting systems through suitable approximation schemes, e.g., perturbative as well as non-perturbative techniques like dynamical mean field theory (DMFT) \cite{Haldar,Bera} and the Random Phase Approximation (RPA).

\begin{acknowledgements}
SB acknowledges support from CRG, SERB (ANRF), DST, India (File No. CRG/2022/001062).
\end{acknowledgements}

\appendix
\renewcommand{\thefigure}{A\arabic{figure}}
\setcounter{figure}{0}

%%%%%%%%%%%%%%%%%%%%%%%%%%%%%%%%%%%%%%%%%%%%%%%%%%%%%%%%%%%%%%%%%%%%%%%%%%%%%%%%%%%%%%%%%%%%%%%%%%%%%%%%%%%%%%%%%%%%%%%%%%%%%%%%%%%%%%%%%%%%%%%%%%%%%%%%%%%

%%%%%%%%%%%%%%%%%%%%%%%%%%%%%%%%%%%%%%%%%%%%%%%%%%%%%%%%%%%%%%%%%%%%%%%%%%%%%%%%%%%%%%%%%%%%%%%%%%%%%%%%%%%%%%%%%%%%%%%%%%%%%%%%%%%%%%%%%%%%%%%%%%%%%%%%%%%

\begin{figure*}[ht!]
\centering
\includegraphics[width=0.8\textwidth]{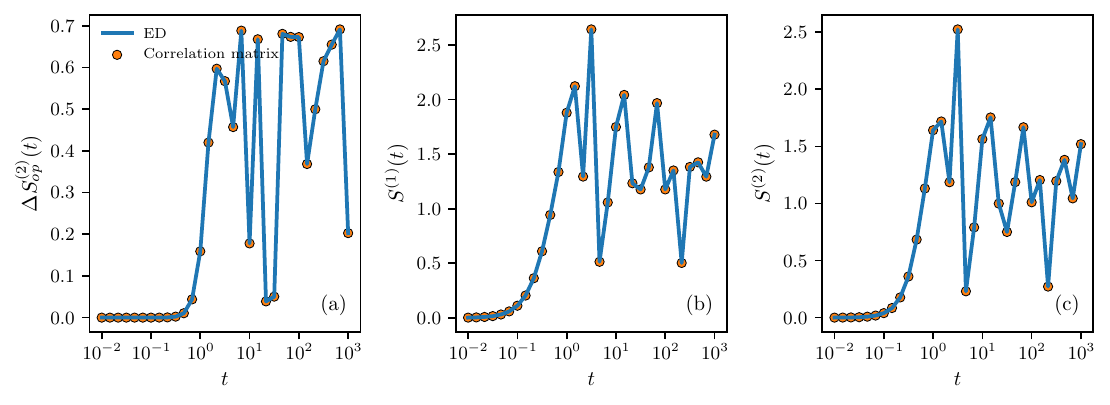}
\caption{Benchmark of (a) operator R\'{e}nyi entropy $S_{op}^{(2)}$ [Eq.\eqref{eq:OperatorRenyi}], (b) von Neumann entanglement entropy $S^{(1)}$ [Eq.\eqref{eq:StateEE}], and (c) second R\'{e}nyi entanglement entropy $S^{(2)}$ [Eq.\eqref{eq:StateRenyiEE}] for $1D$ nearest-neighbor tight-binding chain for $L=8$ with direct numerical many-body exact diagonalization (ED). The time evolutions of $S^{(1)}$ and $S^{(2)}$ have been computed starting from N\'{e}el state.}
\label{fig:AA1D_Benchmark}
\end{figure*}
\section{Initial value of subsystem operator R\'{e}nyi entropy and upper bound on its growth}
\subsection{Proof of $S_{\mathcal{O} A}^{(2)}(t=0) = ( L/2 - 1 ) \ln 2$}
\label{app:S0}

We consider a spin-less fermionic chain of total length \(L\), with Hilbert space $H = H_A \otimes H_B$,
divided into subsystems \(A\) and \(B\) with length $L/2$ (with \(L\) even), so that
$d_A = \dim H_A = d_B = \dim H_B = 2^{L/2}$, and the total Hilbert space dimension $d = d_A d_B = 2^L$. The operator of interest is a local number operator $\mathcal O = \hat{n}_l=\sum_{\{n_{i\neq l}=0,1\}}|n_1,\cdots,n_{l-1},1,\cdots\rangle \langle n_1,\cdots,n_{l-1},1,\cdots|$, at site \(l \in A\) with eigenvalues \(0\) and \(1\). This implies $Z_{\mathcal{O}}=\mathrm{Tr}[\mathcal{O}]=\mathrm{Tr}[\hat{n}_l]=2^{L-1}=d/2$
in Eq.\eqref{eq:DefOperatorRenyi} for $S_{\mathcal{O}A}^{(2)}(0)$. Since $\hat{n}_l$ only has local support in $A$ subsystem, $\mathrm{Tr}_B[\mathcal{O}]=\mathrm{Tr}_B[\hat{n}_l]=d_B\hat{n}_{lA}$, where $\hat{n}_{lA}=\sum_{\{ n_{i\neq l,i\in A}=0,1\}}|n_1,\cdots,n_{l-1},1,\cdots\rangle \langle n_1,\cdots,n_{l-1},1,\cdots|$. As a result, $(\mathrm{Tr}_B[\hat{n}_l])^2=d_B^2\hat{n}_{lA}$ with $\hat{n}_{lA}^2=\hat{n}_{lA}$, and hence $\mathrm{Tr}_A[(\mathrm{Tr}_B[\hat{n}_l])^2]=d_B^2d_A/2$. This leads to $S_{\mathcal{O}A}^{(2)}(0)=\ln(d_A/2)=(L/2-1)\ln{2}$ from Eq.\eqref{eq:DefOperatorRenyi}.

\subsection{Proof of $\Delta S_{\mathcal O A}^{(2)}(t) \le \ln 2$}\label{app:Smaxbound}

The positive semi-definite normalized effective reduced density matrix $\rho_{\mathcal{O}A}(t)=\mathrm{Tr}_B[\mathcal{O}(t)]/Z_{\mathcal{O}}$ defined from the Hermitian operator $\mathcal{O}(t)$ has eigenvalues $\lambda_i\geq 0$ with $\sum_{i=1}^{d_A}\lambda_i=1$. As a result, the subsystem operator R\'{e}nyi entropy $S_{\mathcal{O}A}^{(2)}(t)=-\ln(\sum_i \lambda_i^2)$ from Eq.\eqref{eq:DefOperatorRenyi}. This quantity is upper bounded by minimizing $\sum_i\lambda_i^2$, namely for uniform $\lambda_i$ with $\lambda_i=1/d_A$ for all $i$. 
Consequently,
\begin{equation*}
\sum_{i=1}^{d_A} \lambda_i^2 \ge \sum_{i=1}^{d_A} \left(\frac{1}{d_A}\right)^2 = \frac{1}{d_A}.
\end{equation*}
Thus,
\begin{equation*}
S_{\mathcal O A}^{(2)}(t) \le \ln d_A .
\end{equation*}
% The growth of the operator Rényi entropy is defined as \(\Delta S_{\mathcal O A}^{(2)}(t) = S_{\mathcal O A}^{(2)}(t) - S_{\mathcal O A}^{(2)}(0)\). 
Using this upper bound, we get
\begin{align*}
\Delta S_{\mathcal O A}^{(2)}(t)&=S_{\mathcal O A}^{(2)}(t) - S_{\mathcal O A}^{(2)}(0)
\le \ln d_A - S_{\mathcal O A}^{(2)}(0)\\
&\leq \ln{d_A}-\ln(d_A/2)=\ln{2},
\end{align*}
since $S_{\mathcal{O}A}^{(2)}(0)=\ln(d_A/2)$, as derived in the preceding section.
% Using the result derived in Appendix~\ref{app:S0}, \(S_{\mathcal O A}^{(2)}(0) = \ln(d_A/2)\), we write
% \begin{align*}
% \Delta S_{\mathcal O A}^{(2)}(t)
% &\le \ln d_A - \ln\left( \frac{d_A}{2} \right) \nonumber \\
% &= \ln d_A - (\ln d_A - \ln 2) \nonumber \\
% &= \ln 2.
% \end{align*}
Thus, the growth of the operator Rényi entropy for a local number operator is bounded by \(\ln 2\).

%%%%%%%%%%%%%%%%%%%%%%%%%%%%%%%%%%%%%%%%%%%%%%%%%%%%%%%%%%%%%%%%%%%%%%%%%%%%%%%%%%%%%%%%%%%%%%%%%%%%%%%%%%%%%%%%%%%%%%%%%%%%%%%%%%%%%%%%%%%%%%%%%%%%%%%%%%%

\section{Non-Interacting Green functions and operator R\'{e}nyi and state entanglement entropies}\label{app:NIGF}

For non-interacting systems, to evaluate operator R\'{e}nyi entropy from Eq.\eqref{eq:OperatorRenyi} using the correlation matrix of Eq.\eqref{eq:OpCorrMatrix}], and state entanglement entropies from Eqs.\eqref{eq:StateRenyiEE},\eqref{eq:StateEE} using the correlation matrix of Eq.\eqref{eq:StateCorrMatrix}, we obtain the non-interacting single-particle Green's functions $G_{ij}(ts,t's')$ ($s,s'=\pm$) on the Keldysh contour [Fig.\ref{fig:ctc}]. To this end, we obtain single-particle Green's functions in the basis of the single-particle eigenstates $\{|\alpha\rangle\}$, i.e., $\langle i|\alpha\rangle=\phi_\alpha(i)$ with eigenvalues $\epsilon_\alpha$, of the non-interacting 1D and 2D Aubry-Andr\'{e} models and 2D Anderson model for each disorder realization. Here $|i\rangle$ refers to single-particle site basis state at site $i$. The specific lesser and greater components are given by,
\begin{subequations}
\label{eq:green_functions}
\begin{align*}
    i G_\alpha^{<}(t, t') &= \langle c_\alpha(t+) \bar{c}_\alpha(t'-) \rangle = -n_{\mathrm{F}}(\epsilon_\alpha) e^{-i \epsilon_\alpha (t-t')}, \\
    i G_\alpha^{>}(t, t') &= \langle c_\alpha(t-) \bar{c}_\alpha(t'+) \rangle = [1-n_{\mathrm{F}}(\epsilon_\alpha)] e^{-i \epsilon_\alpha (t-t')}, 
    % \\
    % i G^{\mathbb{T}}(t, t') &= \langle c_\alpha(t+) \bar{c}_\alpha(t'+) \rangle \nonumber \\
    % &= \theta(t-t') i G^{>}(t, t') + \theta(t'-t) i G^{<}(t, t'), \\
    % i G^{\tilde{\mathbb{T}}}(t, t') &= \langle c_\alpha(t-) \bar{c}_\alpha(t'-) \rangle \nonumber \\
    % &= \theta(t'-t) i G^{>}(t, t') + \theta(t-t') i G^{<}(t, t'),
\end{align*}
\end{subequations}
where $n_{\mathrm{F}}(\epsilon_\alpha)$ is the Fermi distribution function. As discussed in the main text, the correlation matrix [Eq.\eqref{eq:OpCorrMatrix}] for the operator R\'{e}nyi entropy and the correlation matrix [Eq.\eqref{eq:StateCorrMatrix}] for the state entanglement entropies are obtained, respectively, using the infinite-temperature Green's functions with  $n_{\mathrm{F}}(\epsilon_\alpha) = 1/2$, and the vacuum Green's functions with $n_{\mathrm{F}}(\epsilon_\alpha) = 0$. The eigen-basis Green functions are transformed to the real-space basis via
\begin{equation*}
    G_{ij}(ts, t's') = \sum_{\alpha} \phi_{\alpha}(i) \phi^*_{\alpha}(j) G_{\alpha}(ts, t's').
    \label{eq:real_space_transform}
\end{equation*}

We benchmark the expressions for the correlation matrices [Eqs.\eqref{eq:OpCorrMatrix},\eqref{eq:StateCorrMatrix}] and the entropies [Eqs.\eqref{eq:OperatorRenyi},\eqref{eq:StateRenyiEE},\eqref{eq:StateEE}], evaluated in terms of single-particle Green's functions [Eq.\eqref{eq:OpCorrMatrix}], with those obtained from the full many-body exact diagonalization (ED) of 1D nearets-neighbor tight-binding chain with $L=8$ sites, as shown in Fig.\ref{fig:AA1D_Benchmark}.
%\section{Benchmark}\label{app:benchmark}

%%%%%%%%%%%%%%%%%%%%%%%%%%%%%%%%%%%%%%%%%%%%%%%%%%%%%%%%%%%%%%%%%

%%%%%%%%%%%%%%%%%%%%%%%%%%%%%%%%%%%%%%%%%%%%%%%%%%%%%%%%%%%%%%%%%%%%%%%%%%%%%%%%%%%%%%%%%%%%%%%%%%%%%%%%%%%%%%%%%%%%%%%%%%%%%%%%%%%%%%%%%%%%%%%%%%%%%%%%%%%
\section{Oprator and entanglement growth for 2D Aubry-Andr\'e model with next-nearest-neighbor hopping $t' = 1/3$}\label{app:AA2D_NNN}
In this section, we present additional results for the 2D Aubry-Andr\'{e} model incorporating a next-nearest-neighbor hopping $t'=1/3$. Figure~\ref{fig:AA2D_NNN_L32} illustrates the time evolution of the von Neumann and the second R\'{e}nyi entanglement entropies for an initial N\'{e}el state at $V=0.25$. As discussed in Sec.\ref{sec:2DAA_TemporalGrowth}, unlike the case with $t'=0$ which exhibits anomalous super-ballistic growth of entanglement entropies for small $V$s, addition of the next-nearest hopping leads more expected behavior with early-time super-ballistic growth followed by ballistic growth. Additionally, the scaling of the saturation times $t_{op},~t_{vN}$ and $t_{RE}$ with system size $L$ for $V=0.0$ and $V=0.25$ are displayed in Fig.~\ref{fig:AA2D_NNN_SAT}. As shown in Table~\ref{tab:AA2D_NNN_SAT}, the exponents for the power-law growth of the time scales with $L$ are consistent with nearly ballistic or weakly super-diffusive operator growth.

\begin{figure}[!htbp]
    \centering
    \includegraphics[width = 0.8\linewidth]{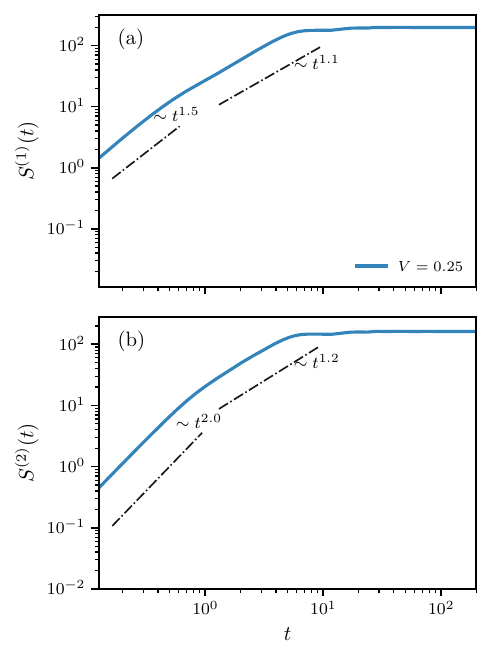}
    \caption{Time evolution of (a) von Neumann, and (b) the second R\'{e}nyi entanglement entropy for an initial N\'{e}el state for the metallic phase ($V=0.25$) for $L=32$ in the 2D Aubry-Andr\'e model with next-nearest- neighbor hopping $t' = 1/3$.}
    \label{fig:AA2D_NNN_L32}
\end{figure}

\begin{figure*}[htbp!]
\centering
\includegraphics[width=0.8\textwidth]{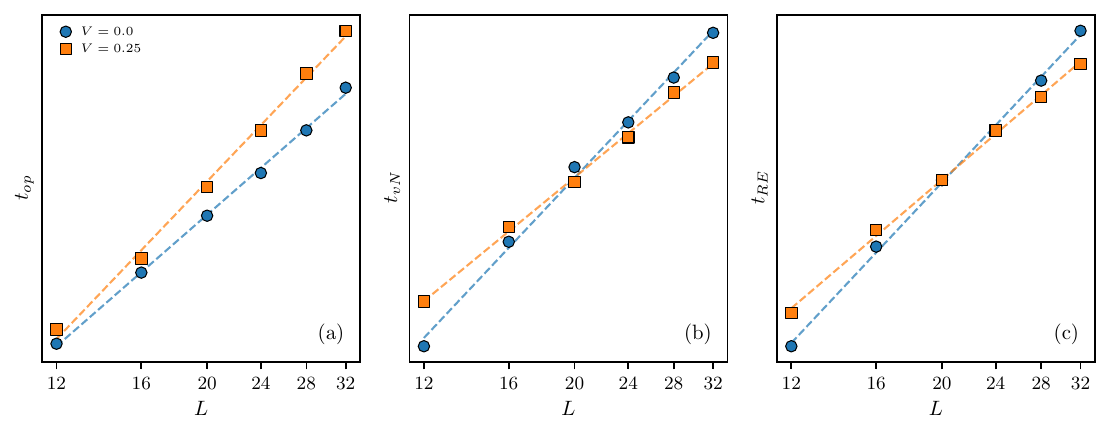}
\caption{Scaling of saturation times, (a) $t_{op}$, (b) $t_{vN}$, (c) $t_{RE}$, associated with the growth of operator R\'{e}nyi entropy and state entanglement entropies, as a function of system-size $L$ in the metallic phase $V=0.0, 0.25$ for the 2D Aubry-Andr\'e model with next-nearest-neighbor hopping ($t'=1/3$).}
\label{fig:AA2D_NNN_SAT}
\end{figure*}

\begin{table}[H]
\centering
\begin{tabular}{|c|c|c|c|}
\hline
V & \textbf{$\alpha_{op}$} & \textbf{$\alpha_{vN}$} & \textbf{$\alpha_{RE}$} \\
\hline
0     & 1.0  & 1.2  & 1.0  \\
0.25  & 1.2  & 0.9  & 0.8  \\
\hline
\end{tabular}
\caption{The exponents $\alpha_{op},~\alpha_{vN},~\alpha_{RE}$ for the power-law scaling of the saturation times $t_{op},~t_{vN},~t_{RE}$ in the metallic phase with next-nearest-neighbor hopping ($t'=1/3$).}
\label{tab:AA2D_NNN_SAT}
\end{table}

%%%%%%%%%%%%%%%%%%%%%%%%%%%%%%%%%%%%%%%%%%%%%%%%%%%%%%%%%%%%%%%%%%%%%%%%%%%%%%%%%%%%%%%%%%%%%%%%%%%%%%%%%%%%%%%%%%%%%%%%%%%%%%%%%%%%%%%%%%%%%%%%%%%%%%%%%%%

\section{Additional results for Schmidt values in 1D and 2D Aubry-Andr\'{e} models}\label{app:Schmidt}

\begin{figure}[htbp!]
    \centering
    \includegraphics[width = 0.9\linewidth]{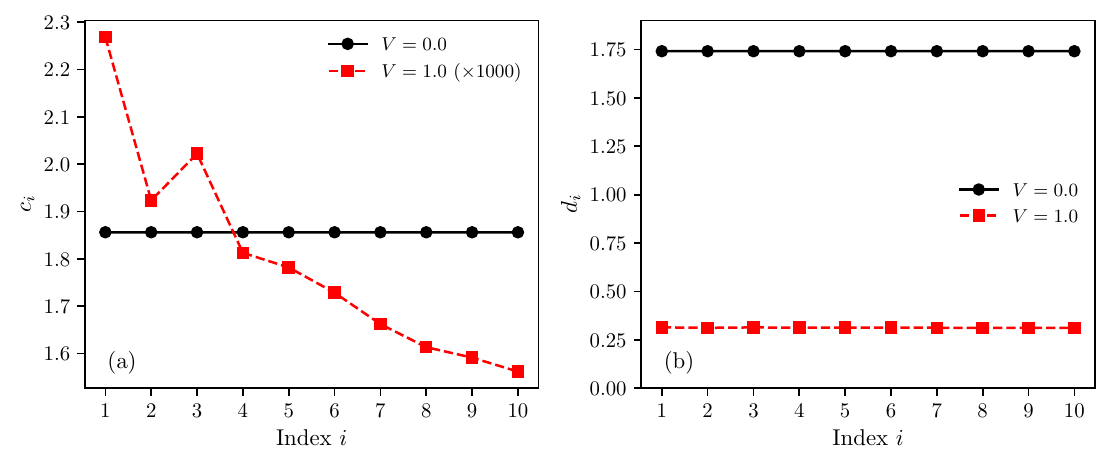}
    \caption{The fit parameters of the ten largest Schmidt values $\Lambda_i(t)$ ($i=1,2,\cdots,10$) for an initial N\'{e}el state in the 1D Aubry-Andr\'{e} model for $V=0.0$ and $V=1.0$.}
    \label{fig:AA1D_Schmidt_Fit}
\end{figure}

\begin{figure}[htbp!]
    \centering
    \includegraphics[width = 0.8\linewidth]{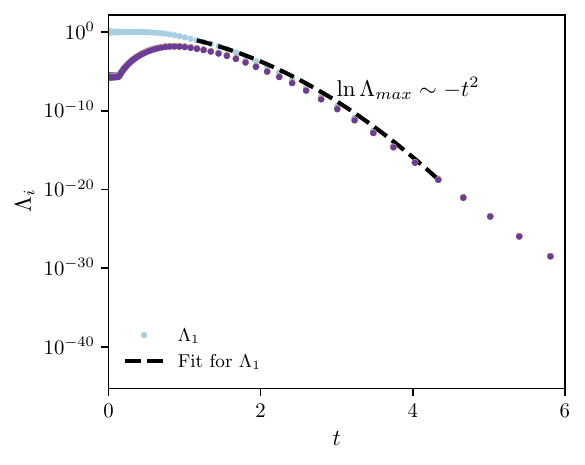}
    \caption{The time evolution of the ten largest Schmidt values $\Lambda_i(t)$ ($i=1,2,\cdots,10$) for an initial N\'{e}el state in the 2D Aubry-Andr\'{e} model for $V=0.25$. The dashed lines indicate fits in the early-to-intermediate-time window with $\ln\Lambda_i\sim -t^2$.}
    \label{fig:AA2D_Schmidt_V0p25}
\end{figure}
The parameters $c_i$ and $d_i$ extracted from the intermediate-time decay, $\Lambda_i=c_i\exp(-d_it^\beta)$ [Fig.\ref{fig:Schmidt_Decay}(a),(b)], of the ten largest many-body Schmidt values are plotted in Figs.~\ref{fig:AA1D_Schmidt_Fit}(a),(b), with the Schmidt indices $i=1,\cdots,10$ for the 1D Aubry-Andr\'{e} model at $V=0$ ($\beta=1$) and $V=1$ ($\beta=0.5$). We observe that the coefficients $c_i$ are suppressed by three orders of magnitude at the nearly diffusive critical point ($V=1$) compared to the clean ballistic limit ($V=0$). Furthermore, while $c_i$ remains nearly constant in the ballistic limit, its magnitude decays with the index $i$ in the diffusive case. In contrast, the coefficients $d_i$ show negligible variation with the Schmidt index $i$, regardless of the transport regime. Consequently, we find no qualitative distinction between the decay profiles of the largest and typical Schmidt values in either cases. Finally, complementing Fig.~\ref{fig:Schmidt_Decay}, we also show the decay of Schmidt values for the 2D Aubry-Andr\'{e} model at $V=0.25$ in Fig.~\ref{fig:AA2D_Schmidt_V0p25}. The intermediate-time decay, $\Lambda_i=c_i\exp(-d_it^2)$, is consistent with the anomalous superballistic behavior seen for small $V$s in the nearest-neighbor 2D Aubry-Andr\'{e} model, as discussed in Sec.\ref{sec:2DAA_TemporalGrowth}.

\clearpage

\bibliography{bibliography.bib}
%\bibliography{RenyiEntropyDMFT}

\end{document}